# Room-temperature near-infrared optical readout of triplet spin states integrated in carbon nanotubes towards quantum sensing.

J. Alejandro de Sousa*, Ilias Vandevenne, Jana Nikolić, Tuur Boonen, Etienne Goovaerts, Sofie Cambré*

*Theory and Spectroscopy of Molecules and Materials (TSM²), Physics Department, University of Antwerp, Belgium*

*Corresponding authors:* JesusAlejandro.DeSousaRodriguez@uantwerpen.be and sofie.cambre@uantwerpen.be

## Abstract

Optically addressable electron-spin qubits that function at room temperature and under ambient conditions are essential for advancing quantum sensing. Solid-state defect platforms offer robust spin–photon interfaces but lack precise defect positioning and electronic tunability. Molecular spin qubits enable bottom-up chemical design yet suffer from limited photostability. Here, we introduce *para*-halophenyl- (-Cl, -Br, and -I) quantum defects in (6,5) single-wall carbon nanotubes that trap triplet excitons with a spin-photon interface that is operable under ambient, room-temperature conditions. These defects were shown to emit in the near-infrared within the biological transparency window, and their placement on the nanotube surface provides nanoscale proximity to target environments. The deep optical traps formed by the defects yield exceptional photostability. Here, we demonstrate optical initialization and readout of these triplet states at room temperature, provide proof-of-principle sensing of paramagnetic molecules, and highlight potential for on-chip detection. Together, these results establish SWCNT defects as a promising new platform for future quantum sensing.

## Introduction

Optically addressable electron spin qubits offer an attractive platform for quantum sensing applications, in particular when operable under ambient conditions at room temperature.[1] Very well known in this perspective are the solid-state platforms such as the nitrogen-vacancy (NV) center in diamond[2,3] or the silicon-vacancy (SiV) center in SiC,[4,5] where a spin-photon interface allows for reading out the spin states optically through optically-detected magnetic resonance (ODMR) resulting in ultrasensitive sensing with single defects of small magnetic fields, temperature, pH, etc.[6–9] In recent years, specifically designed fully-organic molecular electron spin qubits have been synthesized some of which are operable at room temperature, ranging from small molecules such as linear acenes[10,11] with tailorable photoexcited triplet states[12] towards specifically designed ground- or excited-state $S \geq 1$ molecular qudits.[13–17] As opposed to the solid-state platforms, molecular spin qubits provide a controllable bottom-up synthesis that allows for creating low-cost versatile spin qubits with predesigned optical transitions[12], scalability and possibilities for integration in two- or three-dimensional architectures[18] and with operability across different environments and in different configurations (films, solution, solid-state).[10–18] Moreover, low spin-orbit coupling in these organic molecules inherently also promises long spin coherence times.[10,11,13–19] Yet, photochemical long-term stability of such molecular systems remains an issue, often requiring measurements in inert atmosphere conditions or at decreased temperature.[20,21]

While both the solid-state platforms and the molecular quantum bits are intensively investigated, spin defects in low dimensional materials can potentially revolutionize the field of quantum sensing as they can combine the stability and spin-photon interface of the solid-state platforms with nanoscale proximity sensing due to their very small dimensions. Defects in hexagonal boron-nitride[22], and more recently also in boron-nitride nanotubes[23] have emerged as examples of promising platforms for magnetic field sensing and local thermometry even under ambient conditions. However, typically in these low-dimensional materials, defects are either already present after the synthesis or are created through ion implantation[22,23] and thus scalability and control over the position of the defect remain a challenge. Similarly, $NV^-$ defects in nanoscale diamonds are being investigated, reducing the dimensions to bring the defects in closer proximity to their target analytes, but typically resulting in a reduction of their spin coherence times.[24]

In this work, we synthetically create defects in (6,5) single-wall carbon nanotubes (SWCNTs) through *sp³*-functionalization that were previously shown to trap triplet excitons.[25]

Here, we demonstrate that we can address these $S$=1 spin sublevels optically under ambient conditions using ODMR. As was previously shown,[26,27] the spin defects emit within the biological optical transparency window in the near-infrared, which is beneficial for biological sensing, and close to the O-band for optical data communication. They are located at the surface of the one-dimensional SWCNTs giving potential for highly sensitive detection and permitting the design of recognition centers for different analytes.[28] In addition, they are photochemically stable due to the deep optical traps created by the defects on the SWCNTs, which have previously been demonstrated to behave as single-photon sources at room temperature.[27,29] Creation of such defects has shown to enhance photoluminescence quantum yields significantly.[26] Moreover, the presence of only carbon atoms results in low hyperfine coupling ($^{13}C$ abundancy of only 1.1%), which has previously been demonstrated to lead to microsecond-long spin coherence times for electron spins trapped at these defects.[30] While optical read-out of triplet states in *sp³*-defects of SWCNTs and tuning of intersystem crossing through functional group engineering has been demonstrated at low temperature (2.5 K),[25] in this work, we reveal that through fine control of the spin defects, optical initialization and readout of triplet states can be achieved under ambient conditions and at room temperature.

## Results and Discussion

### Room temperature optical readout of defect-functionalized SWCNTs

We synthetically introduced *sp³* defects via diazonium chemistry on chirality-sorted (6,5) SWCNTs wrapped with the polymer PFO-BPY and dispersed in toluene (see Methods and Fig. 1a), following a previously developed procedure.[31] Samples were prepared using 4-halogen-benzenediazonium tetrafluoroborate (halogen: X = Cl, Br, or I). For simplicity, these functionalized SWCNTs are referred to as (6,5)-Cl, (6,5)-Br or (6,5)-I. Their absorption and emission spectra are presented in Fig. 1b (for (6,5)-I) and Fig. 2a (for the other samples). Successful defect incorporation is confirmed by the appearance of a new red-shifted $E_{11}^{*}$ absorption and emission band at around 1150 nm (inset of Fig. 1b and Fig. 1c), consistent with previous reports.[26,32]

Figures 1d-1e present room-temperature half- and full-field ODMR spectra, measured under ambient conditions directly in toluene dispersions, comparing pristine SWCNTs (grey) with a functionalized (6,5)-I sample (purple). Similar spectra are obtained for the other halogen-functionalized SWCNTs (Fig. 2b). As we will demonstrate further, the spectrum is composed of three contributions originating from two triplet systems, i.e. $TE_t$ corresponding to a triplet system trapped at the defect site and $TE_p$ referring to distant pairs of electrons and holes.

Although all measurements reported in the main text were done on (6,5)-X dispersions with PFO-BPY in toluene, similar results can be obtained with (6,5)-X in water-based surfactant dispersions embedded in a silica gel (see Supplementary Fig. 1 in the supporting information (SI)), though with lower signal-to-noise ratio since these experiments require measurements in a small capillary (significantly smaller quantities) due to high MW absorption of deuterated water. Whereas all defect-functionalized samples exhibit clear ODMR signals (typical integration time: 27 minutes full field and 45 min half field), pristine SWCNTs show only a very weak, barely detectable signal at room temperature, the small signal attributed to intrinsic defects in the SWCNTs. Although the ODMR contrasts remain small, of the order of $10^{-4} – 10^{-3}$ depending on laser power (Fig. 2e and 2f), defect density (Fig. 2d) and microwave (MW) power (Supplementary Fig. 3), they can be easily distinguished from the noise, and the best signal-to-noise ratios are of the order of 100 before any processing. Minimal detectable ODMR contrast in our setup is thus of the order of $10^{-6}$.

In our standard ODMR experiment, the sample is placed in the cavity of a continuous-wave X-band EPR spectrometer and excited at the $E_{22}$ transition of the (6,5) SWCNTs with 561 nm laser excitation. Emission is collected after passing through a long-pass 1050 nm filter to isolate the $E_{11}^{*}$ band, although potential phosphorescence and phonon-side band emission may occur in this wavelength range as well. Meanwhile, the sample is irradiated with an on-off modulated microwave (MW) $\boldsymbol{B_1}$ field at 9.44 GHz (typical modulation frequency 630-730 Hz, maximum MW power 360 mW) and the external magnetic field $\boldsymbol{B_0}$ is scanned. Laser excitation initializes a (potentially non-Boltzmann) spin-polarized triplet population. MW-induced transitions between triplet sublevels at specific magnetic fields depolarize this population, that could either result in an increase (positive ODMR) or decrease (negative ODMR) in emission depending on the specific spin dynamics involved. The intensity of the ODMR signal thus depends on the initial steady-state spin polarization, which is determined both by the initialization channels (laser excitation, intersystem crossing) and the depolarization channels (triplet-triplet annihilation, spin relaxation and non-radiative and radiative decay channels). The ODMR contrast is defined as the relative emission change ($\frac{\Delta PL}{PL}$), measured as the lock-in peak-to-peak voltage normalized by the off-resonant direct PL voltage. Supplementary Fig. 2 shows the raw data (i.e. without this normalization) of the ODMR spectra presented in Figs. 1d-1e, with very similar noise level for pristine and grafted SWCNTs. The much larger noise seen for pristine SWCNTs in the latter figure thus originates from the division by the small off-resonant PL value. MW-power dependent ODMR spectra demonstrate no significant saturation up to the highest MW powers, as seen from the close to linear dependence on MW power (Figs. S3-S4).

Fig. 1d and 1e also show simulations (using the spin Hamiltonian parameters optimized as described below) for comparison with the experimental ODMR spectra, consisting of two different triplet systems, which we will further denote as $TE_t$ (referring to trapped excitons at the defects) and $TE_p$ (referring to distant pairs of electrons and holes). For these simulations, we use the Easyspin simulation package in Matlab.[33] Easyspin simulates electron spin resonance (ESR) transition probabilities from a user-defined spin Hamiltonian (see Supplementary Note 4 for the full definition of the spin system). By including also parameters for the spectral line widths and for the degree of orientation of the triplet systems with respect to the magnetic field (random orientation in solution spectra, preferential orientation in films see Supplementary Fig. 5), Easyspin calculates the corresponding ESR spectrum. Since we measure ODMR contrast as opposed to ESR transition probabilities, an additional empirical ODMR efficiency needs to be included, which can be positive or negative. As discussed later, magnetic-field-orientation-dependent ODMR spectra of in-plane aligned SWCNT films (Fig. 4) demonstrate that the negative ODMR contrast originates from triplet states with their Z-axis aligned close-to-parallel with the external magnetic field $\boldsymbol{B_0}$, whereas other orientations yield positive contrast. Note that previous theoretical calculations have demonstrated that the $sp^3$-defects lead to localized triplet states that have their Z-axis close to parallel to the SWCNT axis,[25] thus the negative ODMR contrast can be assigned to SWCNTs that are aligned close to parallel to the magnetic field. Hence, we include an empirical ODMR efficiency, that assigns negative ODMR contrast to those SWCNTs with their long axis aligned close to parallel with the external magnetic field, and positive ODMR contrast otherwise. We integrate this into the Easyspin simulations by first calculating the ESR transition probabilities for each orientation of the triplet systems with the external magnetic field after which they are integrated over all orientations present in the sample with their respective weights and either a positive or negative ODMR efficiency depending on the specific orientation (see Supplementary Note 4).

The positive and negative (blue and cyan) contributions of $TE_t$, are thus simulated with identical spin Hamiltonian parameters assuming they originate from the same triplet system trapped at the defect sites, but with different, positive or negative ODMR efficiencies depending on the orientation of the triplet system with the external magnetic field.

Aside from the ODMR contributions of $TE_t$, presented by the blue and cyan spectra, we also observe a contribution from distant pairs of electrons and holes, yielding an unresolved ($D \cong E \cong 0$) triplet ODMR signal (green spectrum in Fig. 1f; as also reported previously[25,34]), which we will further refer to as $TE_p$.

To extract the spin Hamiltonian parameters, as well as the line widths and the orientational distribution parameters, and systematically explore the parameter space, we designed a fit model, based on these simulations, that numerically optimizes the spin Hamiltonian simulation parameters, allowing to rigorously extract parameter uncertainties. Analysis of angular-dependent spectra of in-plane aligned SWCNT films (Fig. 4) shows that within the experimental accuracy the $g$-tensor of the $TE_t$ triplet system is isotropic, which is in line with the typically low spin-orbit coupling of the atoms in this purely hydrocarbon composite. Importantly, all spectra in this manuscript were fitted simultaneously with the same spin Hamiltonian parameters, which are provided in Supplementary Table 2 in the SI.

Figure 2 shows that the ODMR signals are largely insensitive to which halogen is substituted at the *para* position of the aryl group, giving highly reproducible ODMR spectra for (6,5)-Cl, (6,5)-Br or (6,5)-I at comparable defect densities (Fig. 2b). In contrast, the RT integrated ODMR contrast increases with defect density (Fig. 2d). Since the division by off-resonant PL to obtain the ODMR contrast directly corrects for changes in emission quantum yield with increased defect density, the observed increase in ODMR contrast is consistent with a larger population of trapped triplet excitons contributing to the signal. This trend is opposite to low-temperature observations for (6,5)-$NO_2$ defects,[25] where higher defect densities resulted in reduced integrated ODMR contrasts, most likely due to spin-spin interactions between excitons trapped at different defect sites. This clear difference suggests that different triplet dynamics give rise to the ODMR spectra at room temperature, as will also be further substantiated by laser power and temperature-dependent measurements.

As laser excitation initializes the spin polarization of the triplet states, we investigate how the laser power affects the ODMR contrast (Figs. 2e-f). The corresponding simulations are provided in Supplementary Fig. S7 in the SI. Note that the ODMR contrast is normalized over the value of the off-resonant PL ($\frac{\Delta PL}{PL}$). In a simple scheme where intersystem crossing is responsible for the population of the triplet states, both $\Delta PL$ and $PL$ are expected to be linearly proportional to the laser power in the low-power limit, which would result in a constant ODMR contrast as a function of laser power. Our results show that the off-resonant $PL$ increases linearly with laser power (inset Fig. 2f), while the integrated ODMR contrast increases with laser power and only at the highest powers the constant ODMR contrast with laser power is achieved. Moreover, the positive ODMR contrast approaches this quasi-constant regime much later than the negative ODMR contrast. Such a supra-linear dependence of $\Delta PL$ with laser power can be ascribed to higher-order effects such as the bi-excitonic triplet-triplet annihilation process to determine the population of the triplet states and thus lying at the origin of the observed ODMR contrast. Additionally, the differences between positive and negative ODMR contrast as a function of the laser power indicates different spin polarization dynamics depending on the orientation of the SWCNTs in the external magnetic field. Note that for all the room-temperature measurements, we filled the cavity with a flow of helium gas (1 atm, at room temperature) for increased heat dissipation from the sample during measurements and thus reduced laser-induced heating. Remaining heating effects anyway make our room-temperature observation of ODMR in these systems a conservative claim.

Previous low-temperature studies did not display significant negative ODMR contrasts for (6,5) $sp^3$-functionalized SWCNTs[25], but did display small negative contrasts for pristine (7,5) SWCNTs when the magnetic field was oriented along the SWCNT axis[34], hence we further investigate the temperature dependence of both contrasts. Fig. 3 presents such a temperature-dependent study for both a frozen toluene dispersion and an in-plane aligned film for (6,5)-I, while Supplementary Fig. 10 presents this for the (6,5)-Br sample. For both sample types, at the lowest temperatures, $TE_p$ originating from the distant pairs of electrons and holes (green data points, $\boldsymbol{D} = \boldsymbol{0}$) display a generally decreasing integrated ODMR contrast with increasing temperature, as expected when thermal Boltzmann repopulation dominates the occupation of the triplet states, hence reducing the spin polarization at higher temperature. On the other hand, at slightly higher temperatures around 140-150 K, a plateau is observed. Similarly, the $TE_t$ positive component generally decreases with increasing temperature (dark blue data points) but showing a small increase at around 140-150 K before decreasing further. In stark contrast, the negative component of this triplet system (cyan data points) behaves differently, as its integrated ODMR contrast increases with increasing temperature up to a point of 150 K after which it decreases again. Together with the previous results, this suggests that different spin polarization dynamics play a role when the SWCNTs are aligned parallel to the external magnetic field (see also Fig. 4), as opposed to other orientations. The exact underlying mechanisms are however difficult to disentangle, due to the different processes involved that define this steady-state spin polarization, as mentioned above. Note that, compared to the preferentially planar distribution, the samples with random orientation have a much smaller fraction of the SWCNTs with their long axis close to the external magnetic field. As a result, the negative components are less prominently observed in the frozen dispersions (Fig. 3b) than in the film samples, where the long axis of the SWCNTs and the magnetic field are both aligned in the plane of the film. The plateau observed for the positive ODMR contributions could potentially originate from underlying contributions of the triplet system showing negative contrasts that cannot be resolved with the current model employed. To conclude, two opposite temperature behaviors are observed, one quickly decreasing with temperature, as expected from Boltzmann populated triplet states, while another one increasing with temperature up to ~150 K that displays a non-thermal spin polarization.

It is well known that SWCNTs possess complicated exciton photophysics, with a dark, symmetry-forbidden singlet exciton and several dark triplet states lying below the bright singlet state.[35] Moreover, for $sp^3$-functionalized SWCNTs it was previously observed that a dark singlet state with higher energy than the bright trapped singlet state exists as well.[36] Optical excitation in the higher energy second optical transitions should result in population of these dark states, especially at low temperatures.[37] Magneto-optical experiments have previously shown that applying a magnetic field along the tube axis can brighten the symmetry-forbidden dark state due to splitting of the degeneracy of the K and K' points resulting in intervalley Coulomb mixing, observing a dark-bright exciton splitting of 5-10 meV.[38–40] On the other hand, mixing between bright and dark states was observed for magnetic fields aligned close to parallel to the SWCNT axis, resulting in the observation of additional emission channels.[38] The correspondence with our results in view of the orientation dependence of the magnetic field with respect to the SWCNT axis, and the compatible temperature range of observation (150 K ~12.9 meV), makes this process the most plausible explanation for the negative ODMR contrast found in our work. While the present data do not allow us to establish this mechanism unambiguously, the available experimental evidence is consistent with such an interpretation. According to this picture, this effect becomes increasingly important at thermal energies comparable to the dark-bright exciton splitting, initially leading to a gradual increase of this ODMR contrast with temperature, while at higher temperatures spin relaxation and Boltzmann statistics take over, which corresponds to the observed general decrease with increasing temperature. Note that although laser-induced heating was minimized, a small difference between the cryostat temperature and local nanotube temperature cannot be completely excluded, which would shift the temperature scale slightly upwards so the comparison between the ODMR maximum near 150 K and the relevant excitonic energy level should be interpreted qualitatively. Since the laser power-dependent measurements show that the off-resonant emission changes linearly with laser power, the excitation efficiency of trapped excitons remains constant throughout the measured range. Also intersystem crossing is not expected to depend on laser power, and thus can hardly explain the supra-linear behavior. Therefore, our results seem to point to the spin-dependent recombination channels to be important to achieve this type of spin polarisation. Future investigation will be needed to unravel the exact mechanism.

Finally, Fig. 4 presents the angular dependence of the ODMR spectrum obtained by rotating the plane of the film in the external magnetic field, with the SWCNTs randomly oriented within the film plane (see scheme in Fig. 4a; see also Fig. 3a). While the low-temperature ODMR spectra (Figs. 4c-4f) show no angular dependence, consistent with previous observations for open- and closed-shell $sp^3$ defects[25], this is very prominent in the room-temperature ODMR spectra (Figs. 4b-4e) exhibit a pronounced angular dependence. Negative ODMR features (dark blue color) appear only when the magnetic field is near-to-parallel with the plane of the film, whereas hardly any negative contribution is detected for the perpendicular orientation. Our simulations (see also Supplementary Fig. 13) thus support that when the magnetic field is aligned close-to-parallel to the SWCNT axis, within a Gaussian distribution with width $\sigma_Z = 23.5° \pm 0.6°$, one or more of the mechanisms defining the steady-state spin polarization along the SWCNT long axis is different, as mentioned above.

**Room-temperature SWCNT sensing of open-shell molecules**

Next, we provide a proof-of-principle for the room-temperature sensing of radicals using triplet spin states trapped in the $sp^3$-functionalized SWCNTs. As a model paramagnetic analyte, we selected the stable, free radical TEMPO, while its closed-shell TEMP compound was chosen as a reference (see structures in Supplementary Fig. 18a in the SI). A dispersion of (6,5)-Cl with high defect density (See Supplementary Fig. 15 in SI) and fixed concentration (Optical Density~ 0.84 $cm^{-1}$ at $E_{11}$) was employed as the sensor, while different aliquots of the TEMPO (or TEMP) stock solution were added to obtain varying concentrations. Figures 5a-5b show the ODMR spectra and fitted amplitudes of the positive and negative ODMR contributions as a function of TEMPO concentration. The amplitudes were normalized by the 4 lowest concentration data points. Supplementary Fig. 18 shows the comparison between a second experiment with (6,5)-Br and TEMPO and a reference experiment with the closed-shell TEMP compound, showing no effect on ODMR contrast for the latter. A model fit was superimposed using the following Hill equation: $1 - \frac{1}{1+(K/C)^n}$ with $C$ the concentration, $n = 0.90 \pm 0.03$ the Hill coefficient and a 50% value $K = 1.9 \pm 0.2$ mM. This Hill plot was fitted simultaneously to the three contributions, showing only small differences between them. A clear decrease in ODMR contrast is observed in a dynamic range between 28.5 mM (90% decrease of signal) and 0.1 mM (10% decrease of signal) for TEMPO, while addition of the closed-shell TEMP compound does not change the ODMR contrast (Supplementary Fig. 18). Importantly, the total emission and ratio between $\boldsymbol{E_{11}}$ and $\boldsymbol{E_{11}^{*}}$ emission peaks do not change with TEMPO concentration (Fig. 5c and PL spectra and fits in Supplementary Fig. 16), showing no difference in excitation efficiency or emission is observed in the studied concentration range. Moreover, the decrease in ODMR intensity cannot be explained by changes in the cavity Q-factor induced by increasing TEMPO concentration and stems from the close interaction of the TEMPO radicals with the SWCNT triplet states (Supplementary Figs. 19 - 20). This change in ODMR contrast observed only for the open-shell configuration agrees with a non-thermal population distribution of the triplet spin states, which can be scrambled by spin-spin interactions with the $S = 1/2$ paramagnetic center in TEMPO. It is well known that an increase in the concentration of paramagnetic species directly influences spin relaxation times (as used in relaxometry methods), thereby reducing spin polarization and thus also the ODMR contrast, consistent with previous observations for $NV^-$ centers, and often exploited for quantum sensing.[41] In this system, however, three distinct contributions (with positive and negative ODMR contrasts) can be utilized to validate the response, and the detection fully operates in the near-infrared, making it highly suitable for future (bio)sensing applications.

Finally, we investigated the effect of dissolved oxygen on the ODMR signals. As shown in Supplementary Fig. 19, the presence of dissolved oxygen does not affect the ODMR contrast, indicating that measurements can be reliably performed under ambient conditions.

**Proof-of-principle ODMR detection with a portable microwave antenna**

Finally, a (6,5)-Br film was placed on top of a 1-mm diameter one-loop MW antenna (see schematic Figure 6a) that allows adjusting the MW frequency between 2.8 and 3.2 GHz. This experimental design was chosen to maximize the negative contributions of the ODMR signal, with $\boldsymbol{B_0}$ parallel to the film plane. The light excitation was performed by guiding the 561 nm laser light through the central channel of a fiber bundle, placed just on top of the active antenna area and collecting the light using the 6 peripheral fibers[42] (See Figure 6a). CW-ODMR experiments were performed using amplitude modulation of the MWs, while scanning the external magnetic field to observe the resonance process around $\boldsymbol{g \sim 2}$. Figure 6 shows the raw data (integration time ~60 min) together with smoothed data, showing nicely the two negative contributions to the ODMR spectrum. The low signal-to-noise ratio could be associated with thermal noise generated by the unintentional dissipation of heat in the proximity of the antenna as well as to the limitations of maximum MW power (33dBm) and laser power (50 mW) that available in this simple design. This experiment serves as a first feasibility demonstration for the detection of SWCNT triplet excitons in a simple and compact antenna-based configuration, opening the door to more elaborate on-chip detection schemes, including also pulsed ODMR. The $sp^3$ defects could for instance be implemented in a point of care sensor, or for local detection of magnetic fields, such as those implemented with $NV^-$ centers[43,44] by using a microscope to locate the sensor on the surface. Nevertheless, improvement of the low ODMR contrasts is essential in order to turn this into a technology. Worth noting is the anisotropic behavior of this novel system, with the potential feature of detecting the orientation of local magnetic fields.

## Conclusions

In summary, our results establish $sp^3$-functionalized (6,5) SWCNTs as a novel, robust platform to optically read out triplet spin states at room temperature through ODMR. We uncover an orientation dependence in the sign and magnitude of the ODMR contrast, revealing a distinct spin-polarization mechanism when the magnetic field is aligned parallel to the nanotube axis. This signal is absent at cryogenic temperatures and increases with temperature up to around 150 K, above which it remains persistent at room temperature. From the observed strong dependence of the negative ODMR contrast on magnetic field alignment and temperature, we hypothesize that thermally activated dark–bright exciton dynamics mediates triplet population pathways. We provide a proof-of-principle for quantitative detection of paramagnetic TEMPO molecules through ODMR contrast suppression, validating the sensitivity of SWCNT triplet states to external spin environments. Finally, the successful proof-of-concept implementation of a straightforward antenna-based ODMR setup underscores the future technological potential of these nanohybrids for integrated, NIR-compatible on-chip sensors. Together, these findings open a pathway toward scalable, biocompatible, and ambient-condition quantum sensing architectures based on carbon nanotube exciton spin physics.

## Methods

**Sorting of (6,5) SWCNTs in Toluene:** (6,5) SWCNTs were extracted from CoMoCAT raw material (Sigma Aldrich, SG65i, 0.34 g $L^{-1}$, contains about 40% of (6,5) SWCNTs) by selective wrapping with poly-[(9,9-dioctylfluorenyl-2,7-diyl)-alt-(6,6')-(2,2'-bipyridine)] (PFO-BPy Lot# 22L014A1, Montreal Optoelectronics, MW = 35 kg mol-1, 0.5 g $L^{-1}$) using shear force mixing (Silverson L2/Air) for 96 h, in toluene.[45] Aggregates and impurities were removed by three consecutive centrifugation steps at 25 °C (30 min 4025 g, 4 h 4025 g and 4 h 16215g).

**Sorting of (6,5) SWCNTs in water:** Unsorted CoMoCAT raw material was solubilized in a DOC 1% (w/w) water dispersion using shear force mixing (Silverson L2/Air) for 96 h.

Aggregates and impurities were removed by one centrifugation step at 20 °C (4 h at 16215 g). Sorted (6,5) SWCNTs were obtained by aqueous two-phase extraction using previously reported protocols.[46]

**SWCNT functionalization via diazonium chemistry in organic solvents:** 4-bromobenzenediazonium tetrafluoroborate (BrDZS) (97%) was purchased from Sigma Aldrich, 4-chlorobenzenediazonium tetrafluoroborate (ClDZS) (95%) and 4-iodobenzenediazonium tetrafluoroborate (IDZS) (95%) were purchased from ENAMINE. All salts were used without further purification. Spectroscopic grade toluene and anhydrous acetonitrile were used as well.

The salts showed different reactivities, hence fine control of the diazonium salt concentration and solvent mixture was required to generate samples with similar degree of functionalization, judged by the relative intensity of the $\boldsymbol{E_{11}^{*}}$ peaks in absorption and emission.

Samples with different defect densities of the 4-halogen-phenyl moiety were prepared following the procedure reported by F. J. Berger *et al*.[47] Reactions were performed using a final 80:20 vol-% toluene/acetonitrile mixture or 70:30 vol-% toluene/acetonitrile mixture (for IDZS which has lower solubility) as a dispersive solvent. (6,5) SWCNT concentration was adjusted to an $\boldsymbol{E_{11}}$ absorbance of 0.4 cm$^{-1}$. 18-crown-6 ether 7.6 mM solution in toluene was added to increase the solubility of the diazonium salts in the medium. The diazonium salt was dissolved in acetonitrile, and an appropriate volume of this solution was added to the reaction mixture. Change of the DZS concentration allowed us to generate samples with different defect densities. After running the reaction in the dark for 18 h at room temperature, the functionalized SWCNTs were filtered using a PTFE membrane (Merck Millipore, JVWP, 0.1 µm pore size) and washed with acetonitrile (20 mL). Before each experiment, a part of the SWCNT-coated membrane was cut and added to pure toluene. Short bath sonication (15 min) was used to disperse the tubes.

**SWCNT functionalization via diazonium chemistry in aqueous environment:**

Sorted CoMoCAT, were transferred from a 1% (wt/wt) sodium deoxycholate (DOC) dispersion in $H_2O$ to a dispersion in $D_2O$ by filtration and redispersion using sonication. For the functionalization, the SWCNT concentration was adjusted to an $E_{11}$ absorbance of 0.1 cm$^{-1}$. The final dispersion in $D_2O$ contains 0.92% (wt/wt) of sodium dodecylbenzene sulfonate (SDBS) and 0.08% (wt/wt) DOC.

The diazonium salt BrDZS was dissolved in 1% (wt/wt) SDBS in $D_2O$ and homogenized using 5 min of bath sonication. An appropriate volume of this solution was added to the reaction mixture to achieve a final diazonium salt concentration of 0.07 mM. The reaction was allowed to proceed in the dark for 12 h at room temperature. The DOC concentration was then adjusted to 1.1% to stop the reaction, and the mixture was filtered. To completely remove the diazonium salt, the tubes collected on a cellulose membrane with a 100 kDa molecular weight cutoff were rinsed three times with 5 ml of 1% DOC (wt/wt) in $H_2O$ and 2 mL 2% DOC (wt/wt) in $D_2O$. Finally, the tubes in the filter membrane were dispersed in 2% DOC (wt/wt) in $D_2O$.

**UV-vis absorption spectroscopy (UV-vis):** UV-Vis absorption spectra were acquired with a Cary 5000 spectrophotometer using quartz microcells (Hellma Analytics, QS material) with a path length of 3 mm, and a sample volume of 60 μL. A baseline correction was applied with the pure solvent spectrum.

**Photoluminescence spectroscopy (PL):** PL measurements were performed using an Edinburgh Instruments FLS1000 equipped with a NIR PMT-1700 detector. The samples were excited by the spectrally-filtered output of a continuous-wave Xenon lamp with an excitation bandpass of 6 nm and an emission bandpass of 3 nm. PL spectra were automatically corrected for the instrumental response. Sample concentrations were kept sufficiently low (OD $< 0.3$ in a 3 mm path length cell) to avoid any reabsorption of the emitted light.

**Optically detected magnetic resonance (ODMR) spectroscopy.**

**Using the MW cavity:** The ODMR spectrometer is based on a continuous-wave X-band EPR spectrometer (Bruker ELEXSYS II E500, 9.44 GHz) equipped with a rectangular TE102 resonator (number 9308) with optical access and with an Oxford liquid helium flow cryostat that can be cooled down to 2.5 K. Samples (liquids, 80 μL, or films) were inserted in 4 mm Suprasil® EPR tubes, cut in length (3 cm) and mounted at the end of a 3 mm quartz rod. Samples were resonantly excited with laser light from a 561 nm laser (Cobolt Jive) slightly de-focused on the samples to illuminate the 3x3 mm total sample area through the optical window of the rectangular TE102 cavity. Emission was collected with an InGaAs photodiode (G12180-230A, Hamamatsu, sensitive up to 1550 nm) by light guiding through the quartz rod. The laser excitation was cut with an 830 nm long-pass edge filter. Additionally, a 1050 nm long pass filter was mostly used to cut the (6,5) $E_{11}$ emission and only select the defect emission or triplet phosphorescence. To collect the ODMR spectra, the MWs were modulated on-off (630-730 Hz) and subsequently introduced to a linear power amplifier (KU PA 9001250-2A, KUHNE electronics), resulting in a MW power of up to 360 mW. The sample was located in the position where the electric field of the MWs was zero to avoid any heating effects from the MWs. The on-off modulation was employed for phase-sensitive detection with the internal lock-in amplifier of the EPR spectrometer. The DC component of the detected PL signal was first blocked (EF500 DC Block, <1Hz, Thorlabs) before inserting in the internal lock-in amplifier of the EPR spectrometer.

Room temperature measurements of SWCNTs dispersions were performed under a small, continuous He gas flow (1 atm; at room temperature), serving as a cooling agent with a high heat conductivity to avoid the evaporation of the toluene by laser heating. Any such heating would imply that the nominal sample temperature was higher than room temperature, making our statement that ODMR is measurable at room temperature a conservative lower bound. Temperature-dependent measurements were performed using a liquid helium flow cryostat. To allow for a direct comparison of ODMR intensities, each spectrum was divided by the number of scans and normalized over the total emission coming from the sample (measured before and after each experiment). Toluene dispersions were each diluted to the same concentration (typically in the range of 0.6-1 $cm^{-1}$), when direct comparisons needed to be made. To avoid strong MW absorption when measuring in water-based dispersions at room temperature, small capillaries of 1 cm length and 1 mm internal diameter were filled with a silica gel dispersion of the desired sample. The silica gel was generated by mixing 10% volume Tetramethyl orthosilicate (TMOS, 99%, Sigma Aldrich) with 90% volume of the SWCNTs dispersions (optical density 0.38 $cm^{-1}$) in $D_2O$. The capillary was sealed with cyanoacrylate glue and Teflon to avoid evaporation during measurements.

Similarly as in previous work,[34] for magnetic-field angle-dependent ODMR measurements, thin films were prepared by drop casting 200 μL of nanotubes dispersion with 1 $cm^{-1}$ $E_{11}$ optical density on a glass coverslip precut to an active area of 10x3 mm. They were fixed inside a short Suprasil® quartz tube which was attached to the above-mentioned quartz rod, such that the film sample could be rotated around the axis of the rod, which is perpendicular to the external magnetic field. An angle of 0° corresponds to the $B_1$ field in the plane of the film. For plotting, a data smoothing algorithm using a weighted average of the 3 adjacent points with binomial weights was performed, which was possible as the original ODMR spectra were obtained with sufficient data points (every 0.02 mT).

**Using the MW antenna:** Towards on-chip applications, we implemented a broadband antenna to generate MWs in the 2.8-3.2 GHz frequency range.[48] It consisted of a single 1 mm inner diameter loop on an electronics board with a 50 Ω load. Thin films fabricated as described before were positioned in direct contact with this antenna for maximum impinging of MW power, under ambient conditions. The antenna was driven by a Rohde & Schwartz SMC100A RF signal generator (13 dBm output) and amplified with a 24 dB amplifier (WanTcom, Inc, WBPA2735A) with a saturation output power of 33 dBm, corresponding to a $B_1$ field of approximately 189 μT, estimated from the frequency of Rabi oscillations.[48] The best ODMR signal intensities were obtained using sine-wave amplitude modulation at 730 Hz (max modulation depth).

**Sensing of a paramagnetic center:** A 200 mM mother solution of 2,2,6,6-Tetramethylpiperidine 1-oxyl, 2,2,6,6-Tetramethyl-1-piperidinyloxy, free radical (TEMPO) was prepared in toluene. Diluted solutions in the range of 100 mM to $1x10^{-8}$ mM were used as stocks. Final solutions were prepared by adding 25 μL of TEMPO stock solutions together with 45 μL of (6,5)-Cl dispersion, leading to a final SWCNT concentration of ~ 0.84 $cm^{-1}$ at the $E_{11}$. The reported final concentrations are final concentrations of TEMPO in these solutions. An ODMR spectrum of each solution was recorded in the MW cavity. As a reference, we performed the same experiment with the molecule 2,2,6,6-tetramethylpiperidine (referred to as TEMP), which has a similar structure than TEMPO but a closed-shell configuration.

To verify the cavity Q-factor was not affected by the addition of TEMPO we measured the resonance dip of the microwave reflection upon loading the cavity with 70 μl toluene solutions of TEMPO at different concentrations(Supplementary Figure 19). The resonace dip was fitted with a Lorentzian curve to extract the MW frequency and full width half max (fwhm) of the dip. Finally, we performed a control experiment to rule out the effect of TEMPO in the MW absorption. TEMPO (100 mM) was added to the sample tube filled with the SWCNT dispersion but inside a small capillary to avoid direct contact with the SWCNTs at the molecular level (Supplementary Figure 20) and we compared these spectra with ODMR performed on a mixed TEMPO/SWCNT dispersion comprising the same number of spins (amounting to a final concentration of 22 mM of TEMPO in the dispersion).

**Funding Statement**

J.A.D.S. acknowledges funding from the European Union through a HORIZON-MSCA postdoctoral fellowship C-BRIGHT (N° 101148568), the Research Foundation – Flanders (FWO) for a postdoctoral fellowship (N° 1276727N) and Small Research Grant BOF program of the University of Antwerp (N° FB260019). S.C. acknowledges funding from the Quantera project OPTRIBITS (FWO: G0AHV24N) and from FWO project (G0AAM25N). J.N. acknowledges a grant from Wübben Wissenschaftstiftung GmbH and support form Erasmus program to perform a secondment at UAntwerp. All authors acknowledge FWO for providing infrastructure funding for the ESR instrumentation (I004920N).

**Author Contributions**

J.A.D.S. designed, supervised, performed and analyzed all experiments. I. V. supported the antenna-based measurements. J. N. prepared and performed part of the initial experiments on toluene dispersions. T.B. prepared the samples and performed the measurements in aqueous dispersions. E.G. supervised the experiments and data analysis. S.C. designed, supervised and analyzed all experiments. J.A.D.S. and S.C. wrote the manuscript with support from all authors.

**Competing Interests**

The authors decleare no competing interests.

**Corresponding Authors**

*Sofie Cambré: sofie.cambre@uantwerpen.be

*J. Alejandro de Sousa: JesusAlejandro.DeSousaRodriguez@uantwerpen.be

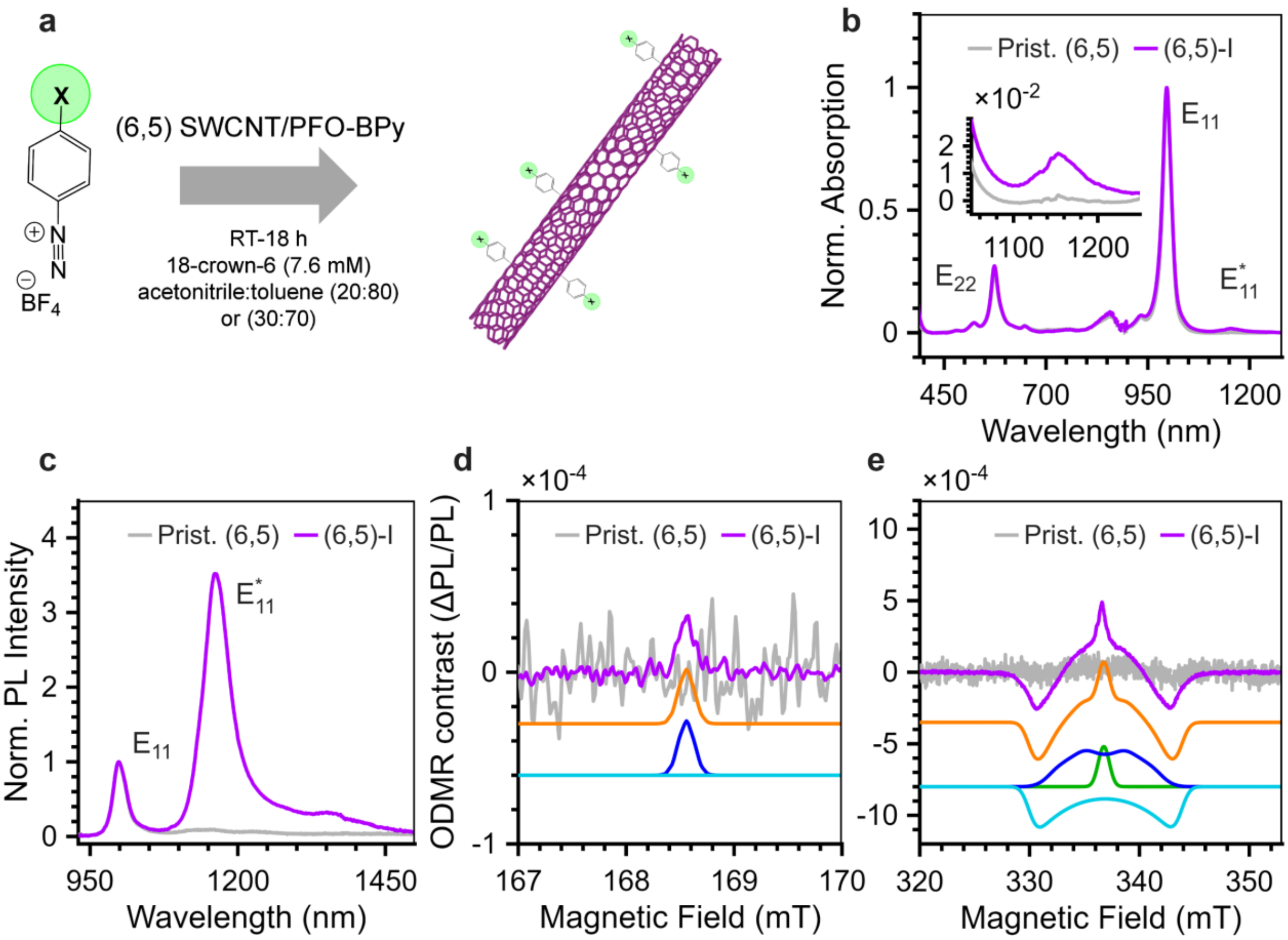

**Figure 1. Room-temperature ODMR of functionalized SWCNTs**. (a) Schematic representation and experimental conditions of the SWCNT functionalization via diazonium chemistry. (b) Normalized absorption spectra of pristine (grey) and (6,5)-I samples (purple). The inset zooms in on the $\boldsymbol{E_{11}^*}$ absorption band. (c) PL spectra obtained at an excitation wavelength of 570 nm and normalized on the $\boldsymbol{E_{11}}$ emission peak. (d) Half-field and (e) full-field ODMR spectra of pristine (grey) and (6,5)-I (purple) samples at room temperature (~ 298 K), with laser excitation at 561 nm, a laser power of 20 mW, a microwave power of 360 mW, a modulation frequency of 630 Hz and both detected by passing the emission through a 1050 nm long-pass filter. The larger noise for the pristine SWCNTs originates from division by a smaller off-resonant PL signal – see Supplementary Fig. 2 for unnormalized spectra. The corresponding simulations are provided in orange as a sum of the green ($TE_p$, $D$=0), blue ($TE_t$, positive contrast) and cyan ($TE_t$, negative contrast) spectra (see main text for explanation). Data presented here are from a single sample. Similar behavior was observed for other functional groups in Fig.2. Ratio's of $\boldsymbol{E_{11}^*}$ versus $\boldsymbol{E_{11}}$ in absorption and PL spectra are provided in Supplementary Table 1.

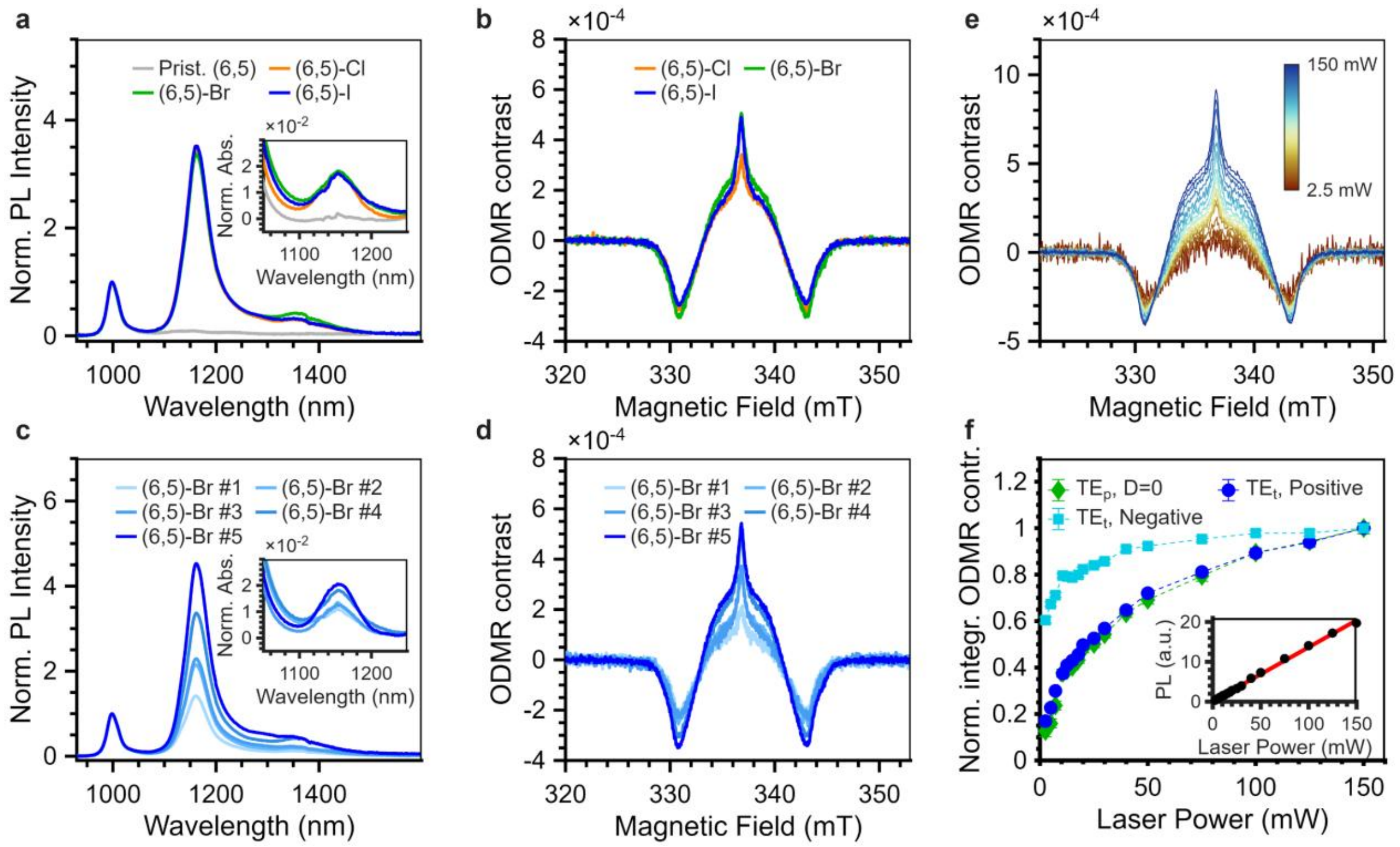

**Figure 2. Dependence on defect density, halogen atom and laser power.** (a) Normalized (to the $E_{11}$ maximum ) PL spectra of pristine (6,5) and functionalized samples (6,5)-X with similar defect density and different functional groups, where X= Cl, Br, I, excitation wavelength 570 nm. The inset shows the corresponding absorption spectra zoomed in on the $E_{11}^{*}$ peak. (b) Full-field ODMR spectra (ODMR contrast = ΔPL/PL) of functionalized samples with similar defect density but different functional groups at room temperature, using laser excitation at 561 nm and a laser power of 20 mW. (c) Normalized (to the $E_{11}$ maximum ) PL spectra of (6,5)-Br samples with different defect density, excitation wavelength 570 nm, and with the corresponding absorption spectra zoomed in on the $E_{11}^{*}$ peak in the inset. (d) Full-field ODMR spectra of (6,5)-Br samples with different defect density at room temperature, using laser excitation at 561 nm and a laser power of 20 mW. (e) Laser power dependence of full-field ODMR spectra for (6,5)-I sample at room temperature, using laser excitation at 561 nm. (f) Integrated ODMR contrast obtained by fitting all the ODMR spectra of panel (e) simultaneously, with the model as described in the text composed of three contributions (with similar color coding as in Figure 1). Corresponding fits are provided in Supplementary Fig. 7. Error bars are 1-$\sigma$ error bars on the fitted amplitudes. The inset displays the perfect linear dependence of the off-resonant PL on the laser power. All ODMR measurements were performed with a modulation frequency of 630 Hz and by passing the emission of the functionalized sample through a 1050 nm long-pass filter. The data presented is from a single measurement on the same sample for all laser powers.

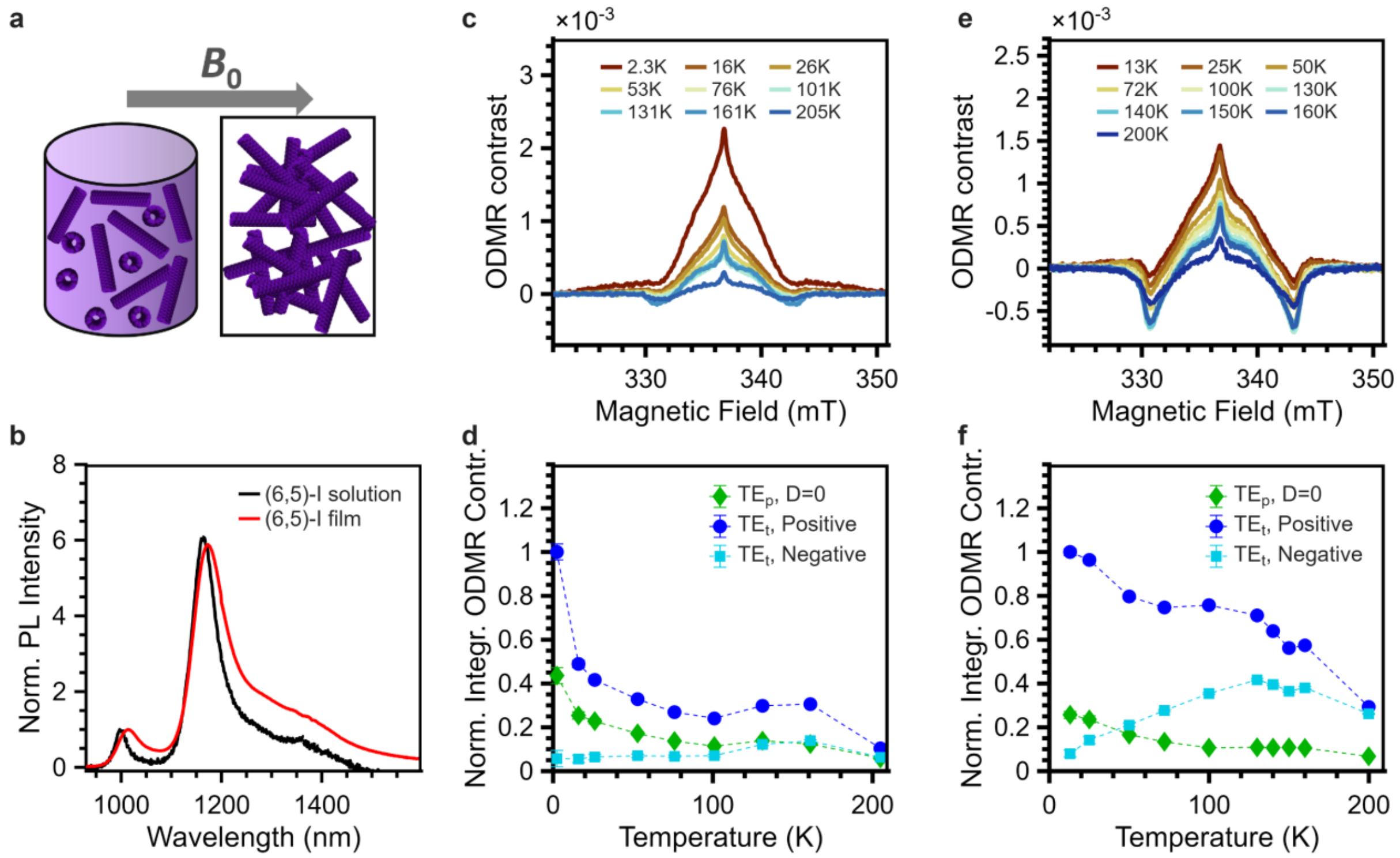

**Figure 3. Temperature-dependent ODMR contrasts** (a) Schematics of a randomly oriented solution and an in-plane aligned SWCNT film inside the external magnetic field, with $\boldsymbol{B_0}$ aligned in the plane of the film. (b) Normalized (to the $E_{11}$ maximum) PL spectra of (6,5)-I samples in solution or in film, where the broadening in the film originates from bundle formation. (c) Full-field ODMR spectra of (6,5)-I frozen solution samples as a function of temperature. (d) Normalized integrated ODMR contrasts from the three different components with which the ODMR spectra of the frozen solutions were fitted, with similar color coding as in Fig. 1. Fits are provided in Supplementary Fig. 8. Normalization was done on the highest value of the three curves. (e) Full-field ODMR spectra of (6,5)-I in-plane aligned film samples as a function of temperature. (f) Normalized integrated ODMR contrasts from the three different components with which the ODMR spectra of the film samples were fitted, with similar color coding as in Figure 1. Fits are provided in Supplementary Fig. 9. Error bars are 1-$\sigma$ error bars on the fitted amplitudes. All ODMR measurements were performed with a MW power of 300 mW and a laser power of 100 mW and by passing the emission of the functionalized sample through a 1050 nm long-pass filter. The data presented is from a single set of measurements taken from low to high temperatures. Temperature-dependent ODMR spectra for (6,5)-Br sample are presented in Supplementary Fig. 10-12.

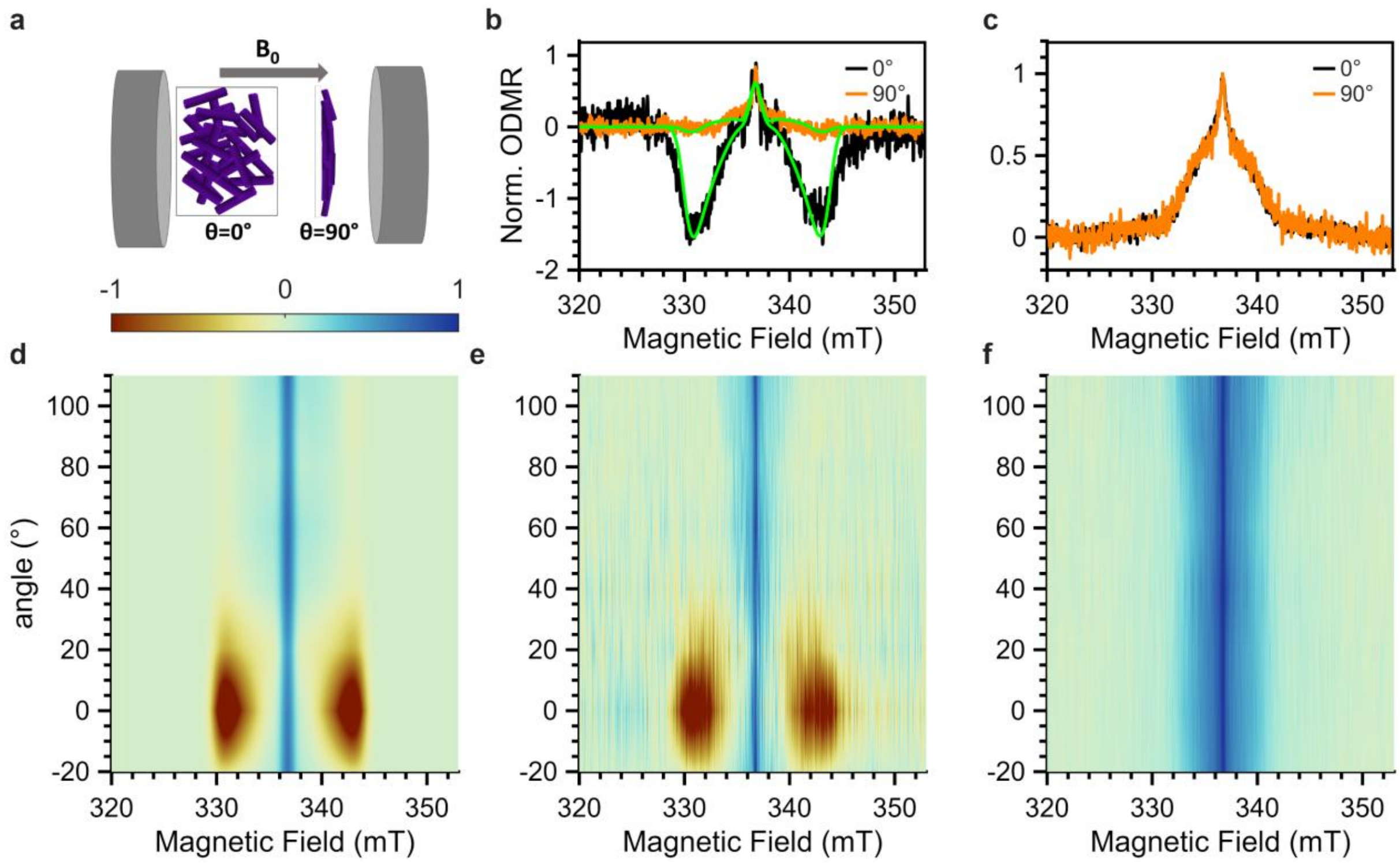

**Figure 4. Angular-dependent ODMR spectra** of in-plane aligned (6,5)-Br SWCNT films (a) schematic definition of the angles of the film plane with the magnetic field, (b) ODMR spectra at room temperature for two discrete angles with superimposed simulations in green. All ODMR spectra in this figure are normalized on the fitted amplitude of the $D = 0$ peak. (c) ODMR spectra at 2.5 K for two discrete angles, (d) Fits of the angular-dependent room-temperature ODMR spectra (see Supplementary Fig. 13 for individual angles), (e) angular-dependent room-temperature ODMR spectra and (f) low-temperature (2.5 K) angular-dependent ODMR spectra. The spectra were measured with a MW power of 300mW, using a long pass filter of 1050 nm, a modulation frequency of 630 Hz, excitation wavelength 561 nm and laser power of 50 mW. The data presented are from a single sample and set of measurements. The angular dependence of (6,5)-I films shows the same behavior (see Supplementary Fig. 14).

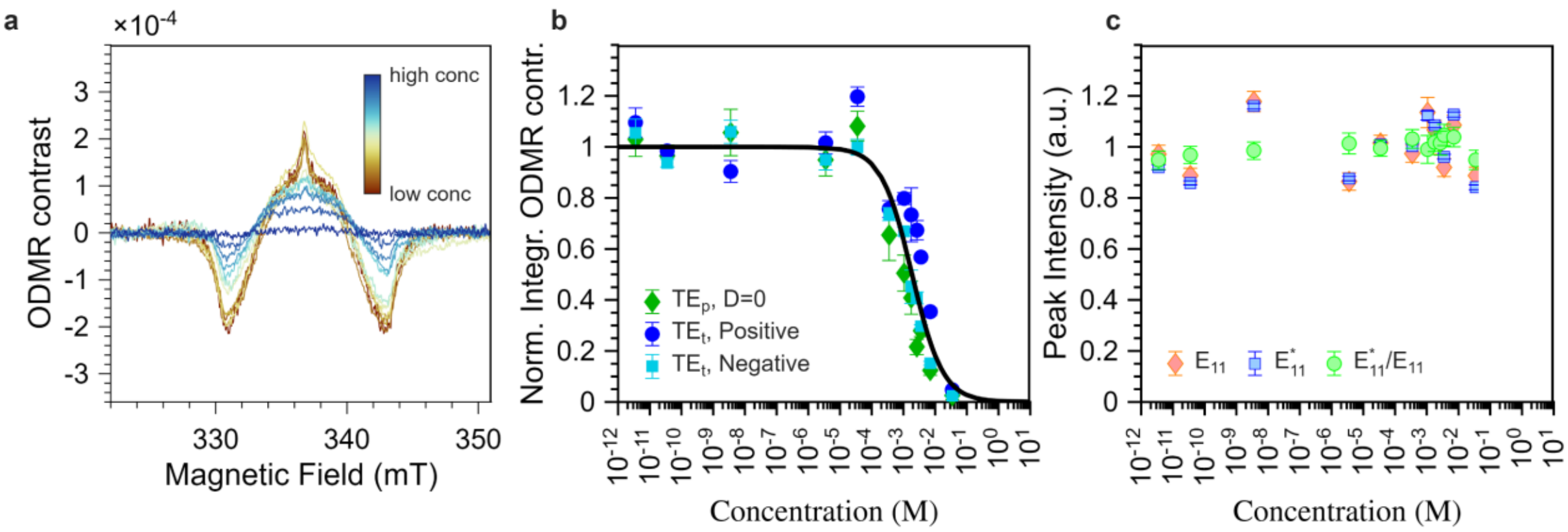

**Figure 5. Sensing of the TEMPO free radical** (a) RT ODMR spectra of (6,5)-Cl sample as a function of added TEMPO concentration showing a general decrease of the ODMR spectrum with increasing TEMPO concentration, using laser excitation at 561 nm and a laser power of 12 mW. and a MW power of 300 mW. (b) Normalized fitted integrated ODMR contrasts (Supplementary Fig. 12 for fits) of the three different components (each normalized on the first 4 data points) are fitted simultaneously with a Hill model (solid line). Error bars represent 1-sigma fitting errors and each data point represents a measurement on an independent sample. (c) Fitted amplitudes of the $\boldsymbol{E_{11}}$ and $\boldsymbol{E_{11}^{*}}$ peaks and their ratio normalized over the mean of all data points, as a function of the TEMPO concentration showing hardly any influence on the SWCNT emission. The reported TEMPO concentrations are final concentrations after mixing the stock solutions. Comparison with the closed-shell TEMP molecule is provided in Supplementary Fig. 18. Reproducible behavior is shown in Supplementary Fig. 18 for (6,5)-Br sample.

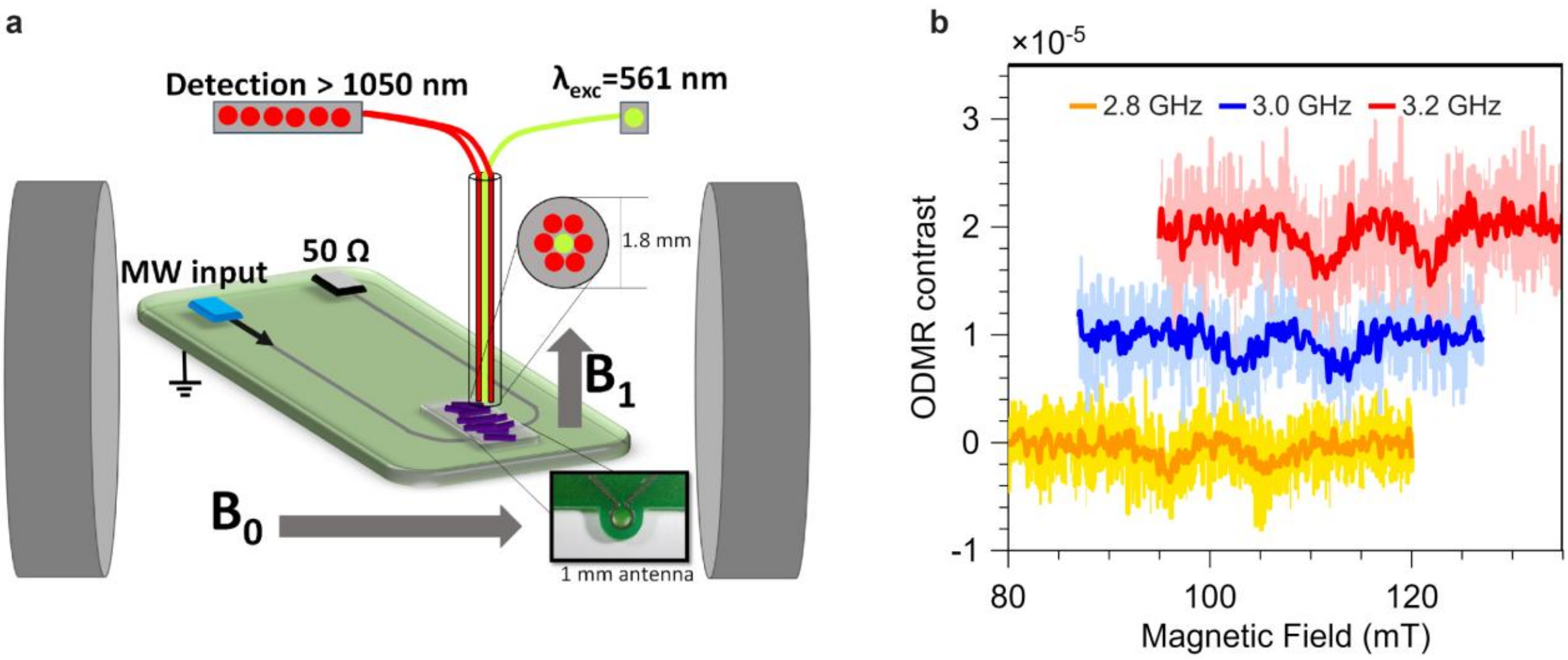

**Figure 6. Prototype on-chip detection** (a) Schematic illustration of an ODMR setup based on an optical fiber bundle for excitation (inner fiber) and collection/detection (outer 6 fibers) of the emission of the SWCNT film placed in close proximity of the 1-mm single-loop MW antenna (photograph as inset). (b)

Measured RT ODMR spectra using this MW antenna operated at three different MW frequencies. The data presented are from the same sample.

**Supplementary Information for:**

**Room-temperature near-infrared optical readout of triplet spin states integrated in carbon nanotubes towards quantum sensing.**

J. Alejandro de Sousa*, Ilias Vandevenne, Jana Nikolić, Tuur Boonen, Etienne Goovaerts, Sofie Cambré*

*Theory and Spectroscopy of Molecules and Materials (TSM$^2$), Physics Department, University of Antwerp, Belgium*

*Corresponding authors:* JesusAlejandro.DeSousaRodriguez@uantwerpen.be and sofie.cambre@uantwerpen.be

# Table of Contents

## Supplementary Note 1: ODMR spectra in water-based dispersions.

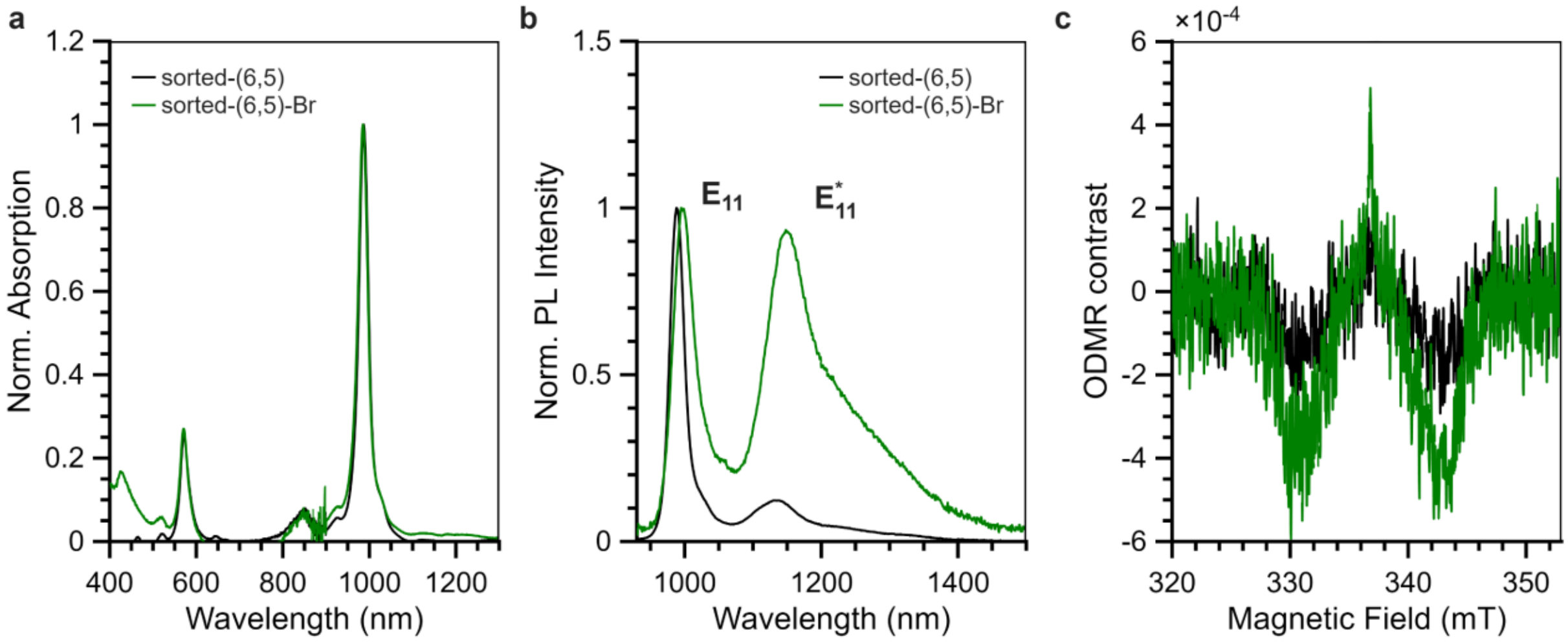

**Supplementary Figure 1**. (a) Background-subtracted absorption spectra of a (6,5) and (6,5)-Br SWCNT surfactant dispersion in $D_2O$, with the background modelled as a $\mathbf{1/\lambda}$ dependence. (b) PL spectra of a silica gel dispersion of sorted (6,5) (black) and (6,5)-Br (green) wrapped with DOC obtained at an excitation wavelength of 570 nm and normalized on the $\boldsymbol{E_{11}}$ emission peak. (c) Full-field ODMR spectra of pristine (black) and (6,5)-Br (green) SWCNTs wrapped with DOC and embedded in a silica gel matrix at room temperature measured in a small capillary, with laser excitation at 561 nm, a laser power of 50 mW, a microwave power of 360 mW, a modulation frequency of 630 Hz and by passing the emission through a 1050 nm long-pass filter. Note that the pristine samples already incorporates intrinsic defects that give rise to a small ODMR signal.

**Supplementary Table 1:** Ratio between E11* peak and E11 in absorption and Emission of samples represented in Figure 1 and Figure 2.

| Sample | $E_{11}^*/E_{11}$ Abs | $E_{11}^*/E_{11}$ PL |
|---|---|---|
| **Pristine** | 0.002 | 0.096 |
| **(6,5)-Cl** | 0.018 | 3.452 |
| **(6,5)-Br** | 0.018 | 3.367 |
| **(6,5)-I** | 0.017 | 3.521 |
| **(6,5)-Br #1** | 0.011 | 1.421 |
| **(6,5)-Br #2** | 0.014 | 2.153 |
| **(6,5)-Br #3** | 0.013 | 2.293 |
| **(6,5)-Br #4** | 0.018 | 3.367 |
| **(6,5)-Br #5** | 0.021 | 4.531 |

## Supplementary Note 2: ODMR spectra before division by off-resonant PL

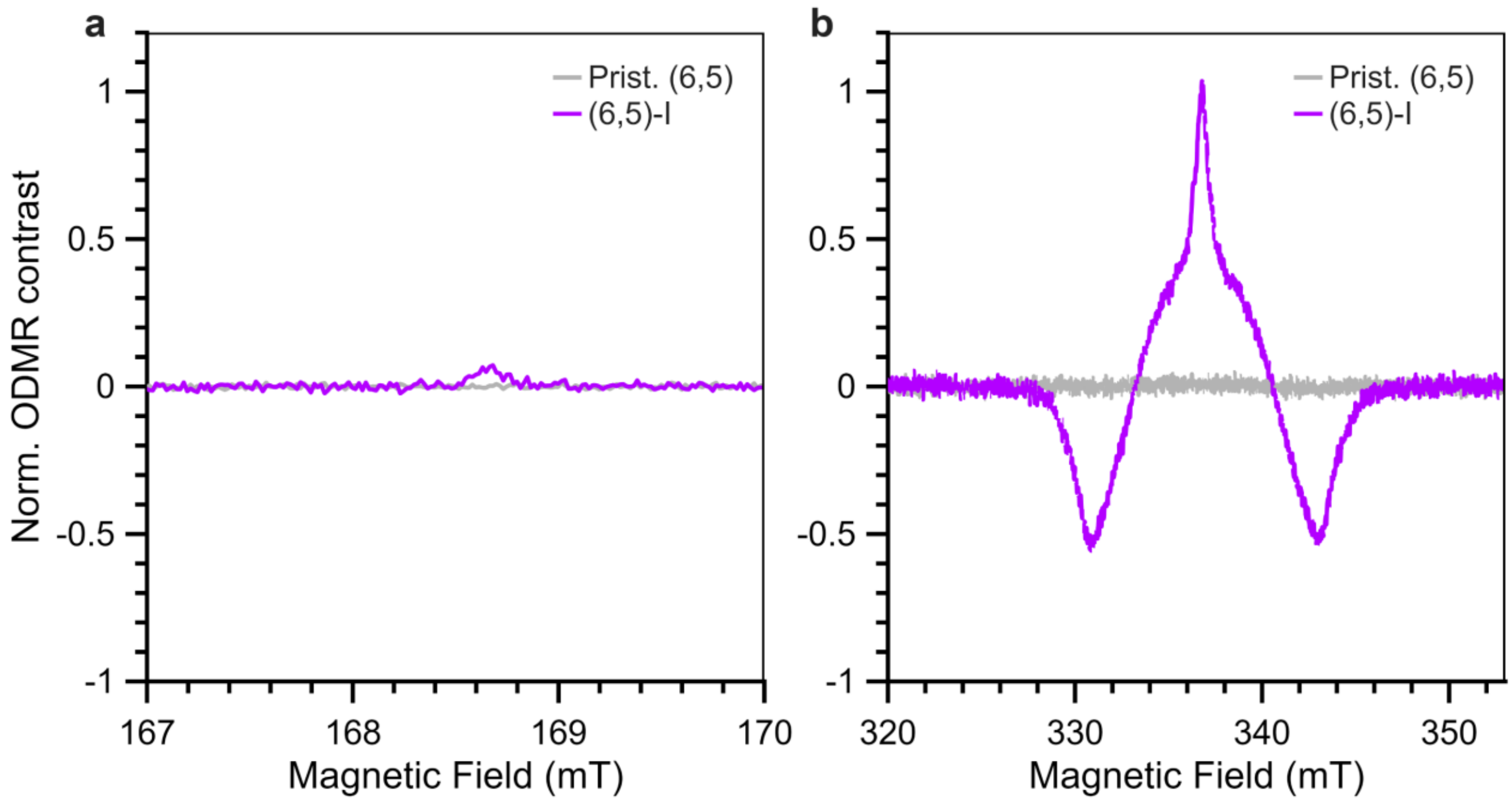

**Supplementary Figure 2.** (a) Half-field and (b) full-field raw ODMR spectra shown in Figure 1 in the main text for the same pristine and (6,5)-I spectra, plotted before division by the off-resonant PL and before smoothing, normalized to the peak of the full-field spectrum for the (6,5)-I sample.

## Supplementary Note 3: MW power dependent ODMR spectra

For ODMR, the signal depends on a change in spin state populations, and thus one expects at low power, far from saturation, that the behavior with microwave power is linear. For increased MW power, the ODMR contrast should reach a constant value, as one detects the difference between PL at initial spin polarization minus the one where populations are nearly equalized. Supplementary Fig. 3 shows this MW dependence for the different spectral contributions each showing weak saturation for the highest MW powers, while Supplementary Fig. 4 shows the fits of the MW-dependent ODMR spectra. While the negative $TE_t$ signal and the $TE_p$ D=0 signal increase similarly with laser power (factor 2.33 and 2.28 in integrated ODMR contrast respectively for highest versus lowest measured powers), the positive $TE_t$ signal increases more steadily (factor 3.23) resulting in a different ratio of the three components depending on MW power (important for comparison with the antenna-based experiments).

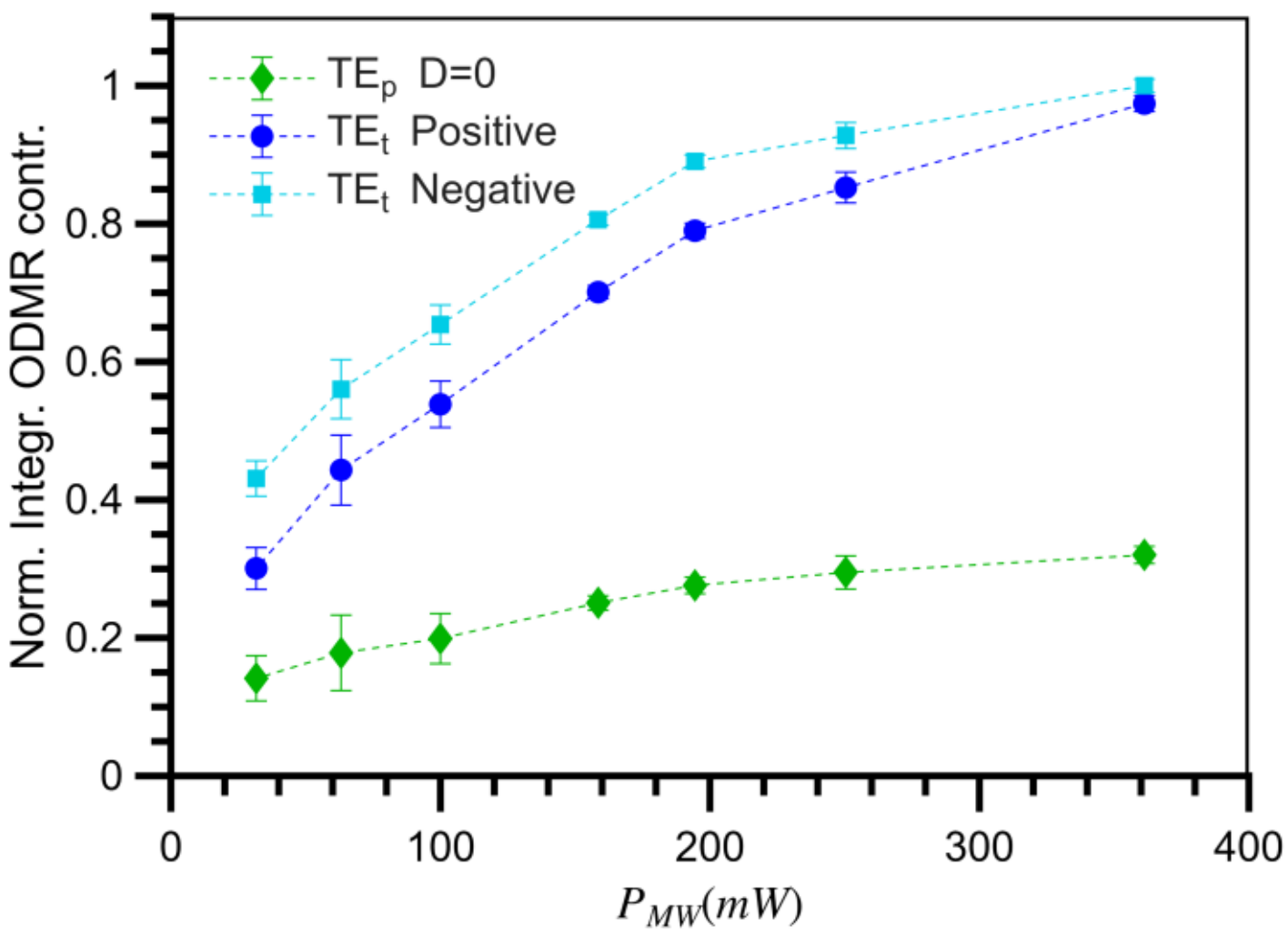

**Supplementary Figure 3.** MW power dependence of the different contributions showing only weak saturation at the highest MW powers as seen from the quasi-linear dependence up to approximately 200 mW.

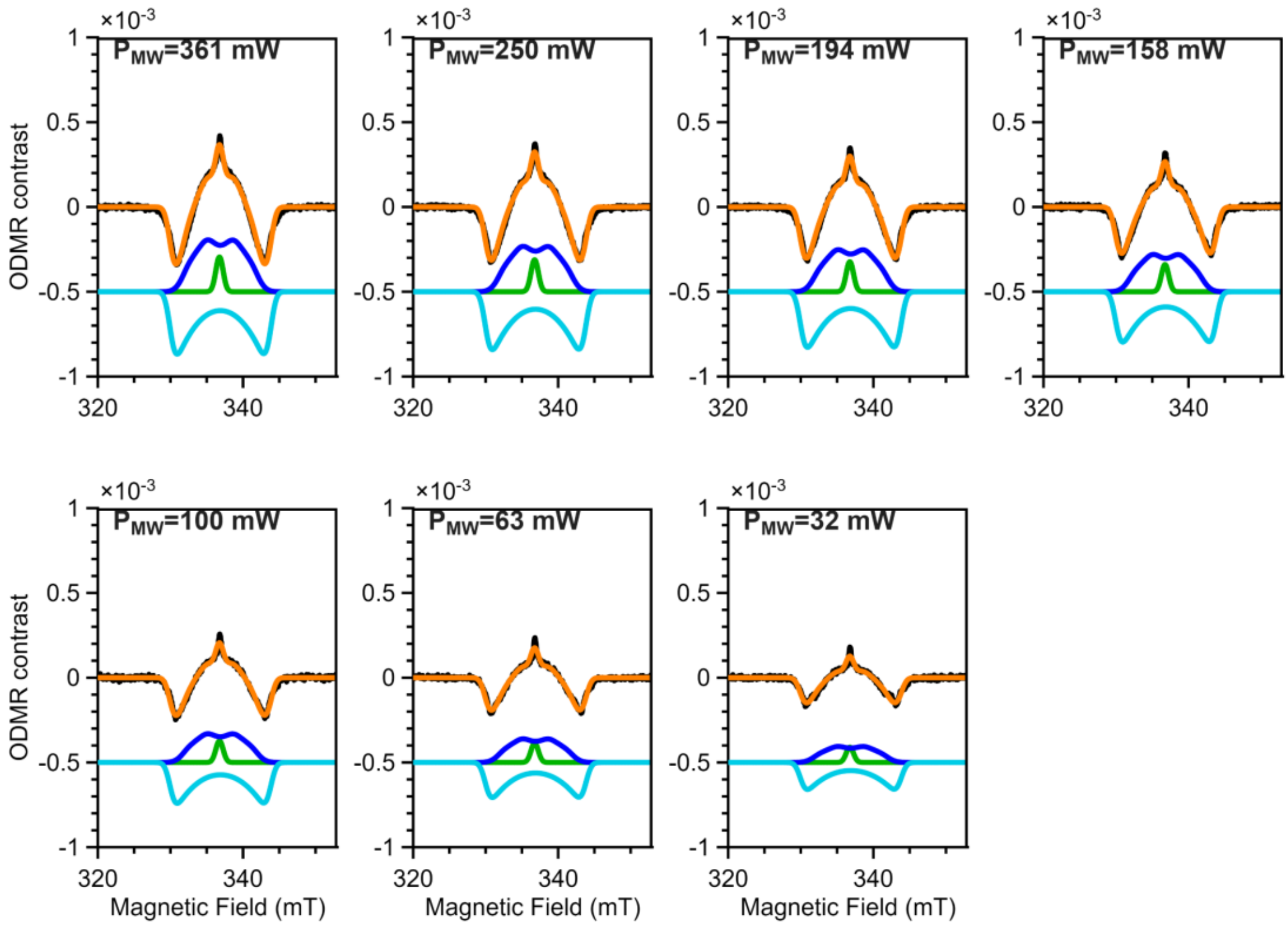

**Supplementary Figure 4**. Simulation of the full-field ODMR spectra in the MW power-dependent measurements, including the 3 components as described in Figure 1 and the main text. Experimental data are shown in black, fits in orange, and components in green ($TE_p$, $D = 0$), dark blue ($TE_t$, positive signals) and cyan ($TE_t$, negative signals).

## Supplementary Note 4: ODMR simulations

**Spin Hamiltonian model for the simulations**: We used the Easyspin simulation package[1], which calculates electron spin resonance (ESR) transition probabilities for different magnetic field orientations and integrates the spectrum over a specific orientational distribution function.

For a triplet system, this spin Hamiltonian is defined by the electron Zeeman and zero-field splitting terms:

$$H = H_{EZ} + H_{ZFS} = \mu_B(g_x B_x \hat{S}_x + g_y B_y \hat{S}_y + g_z B_z \hat{S}_z) + D\left[\hat{S}_z^2 - \frac{1}{3}S(S+1)\right] + E(\hat{S}_x^2 - \hat{S}_y^2))$$

with $\boldsymbol{g}_{x,y,z}$ representing the *g*-tensor parameters, *D* and *E* the zero-field parameters, *S*=1 the total spin quantum number and $\hat{\boldsymbol{S}}_{x,y,z}$ the Pauli spin matrices. Aside from the *g*- and zero-field parameters, the line width of the spectrum and the parameters of the above-mentioned orientational distribution function are also parameters that can be numerically optimized. For the frozen or room-temperature spectra of the dispersions, the orientational distribution function is isotropic, meaning that each orientation has the same probability. For the film samples, where SWCNTs are preferentially oriented with their long axis close to parallel to the plane of the film, we employ a distribution function in which the Z-axis of the triplet system, which was shown to correspond to the SWCNT long-axis from DFT calculations,[2] is preferentially aligned perpendicular to the sample Z-axis (which we define to be normal to the film plane), but randomly oriented with respect to the sample X- and Y-axis, and afterwards the angle of the magnetic field with this sample is changed depending on the experimental conditions, as shown in Supplementary Fig. 5.

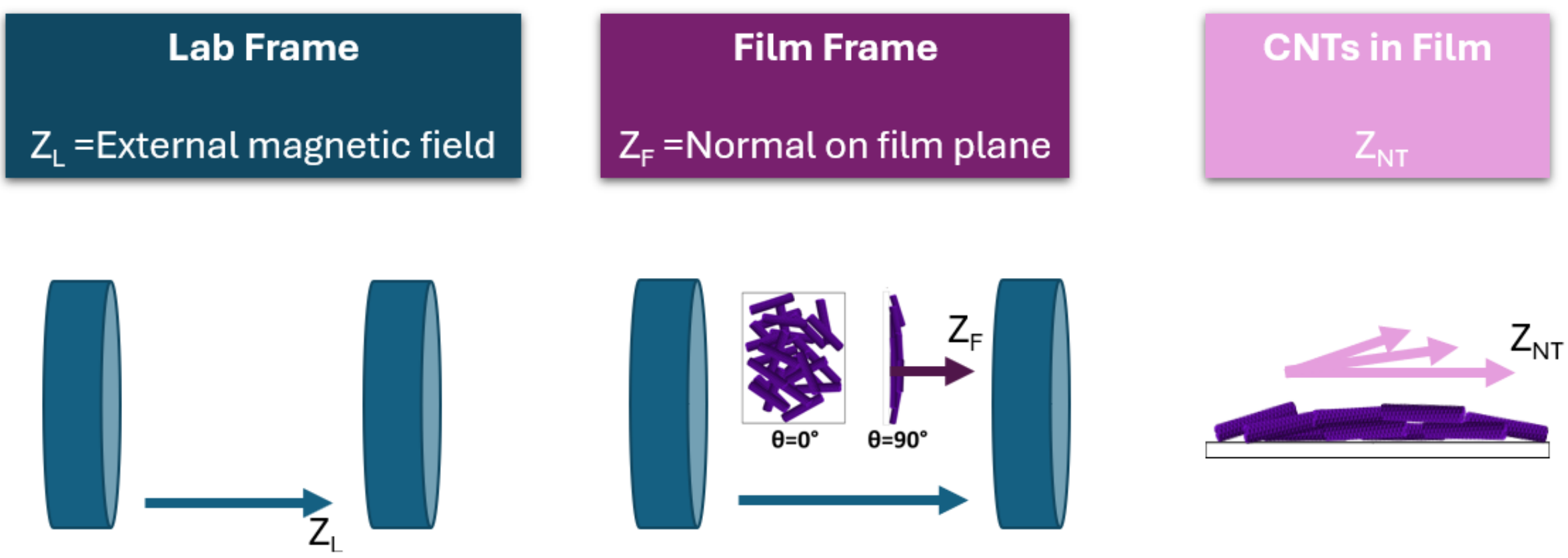

**Supplementary Figure 5**. Schematic representation of the distribution of the CNTs in the plane of the film and the orientation of the film in the lab frame.

We employ a superposition of two different triplet systems, $TE_t$ and $TE_p$. While $TE_t$ displays an isotropic g-tensor and measurable rhombic zero-field splitting parameters, $TE_p$ displays unresolved zero-field splitting parameters, i.e. D≅ E≅ 0 (see Supplementary Table 2 for the optimized parameters). The latter spectrum originates from weak overlap between electron and hole spin density distributions, and is typically observed in SWCNT ODMR spectra, [2,3] while the first one is typically observed for triplet states trapped in $sp^3$-defects.[2] Since the second triplet exciton has unresolved zero-field parameters, and displays a single Gaussian line shape, it can be most easily simulated by an effective S=1/2 system. Supplementary Figure 6 demonstrates that such a simulation is in agreement also with an S=1 system with negligible zero-field splitting but requiring the D and E values not to be incorporated as additional parameters in the simulation. Throughout the manuscript, this contribution is always displayed in green.

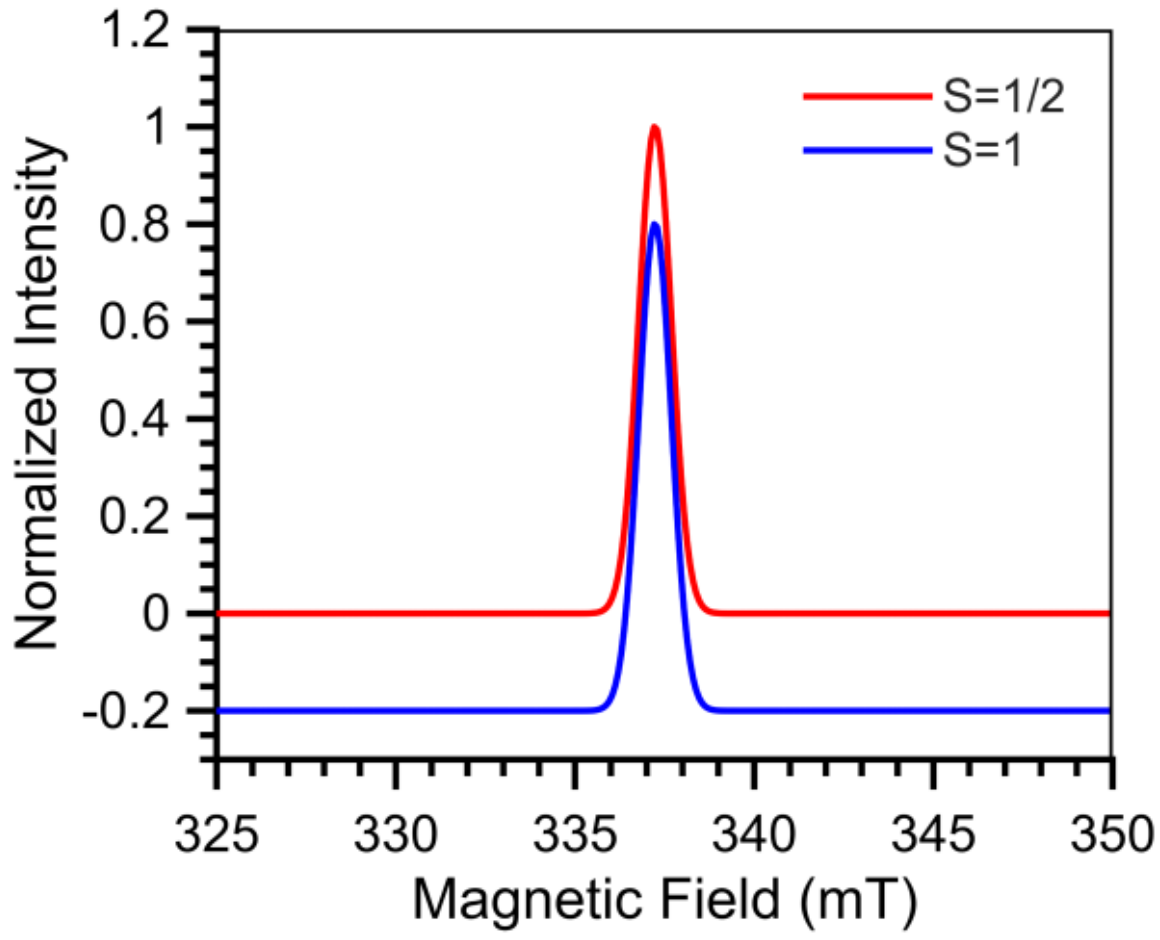

**Supplementary Figure 6** Comparison between a simulation with an *S*=1/2 or *S*=1 spin system for the $TE_p$ component, the latter with unresolved zero-field splitting showing that both give the same Gaussian line shape and thus are interchangeable.

**Adaptation for ODMR efficiency**: Easyspin is designed for simulating ESR transition probabilities, and not ODMR, and therefore does not take into account the efficiency by which spin transitions contribute to the change in emission intensity. In particular, it does not directly allow us to simulate a spectrum with both positive and negative ODMR signals as obtained in this and previous work[3] on SWCNTs. Hence an overall ODMR efficiency, depending on the specific orientation of the SWCNTs with the magnetic field axis, needs to be incorporated.

The ODMR spectra reported herein consist of both positive and negative contributions due to the increase or decrease of the PL upon magnetic resonance, respectively. As we demonstrate in the main text, this change in the sign of the ODMR contrast is associated with different orientations of the long-axis of the SWCNTs with respect to the externally applied magnetic field. Indeed, angular dependent measurements were performed of in-plane aligned SWCNTs, showing negative contrast only when the externally applied magnetic field is close to parallel to the film plane, and thus close to parallel to the long-axis of part of the SWCNTs.

For the negative ODMR contribution of the $TE_t$ spectrum, first an ESR spectrum with a Gaussian orientational distribution function with maximum for SWCNTs aligned with the laboratory $Z_L$-axis (*i.e.* $\boldsymbol{B_0}$) is simulated. The width of this Gaussian distribution ($\boldsymbol{\sigma_Z}$) **is** also a parameter in the Easyspin simulation program. We then assign mathematically a negative amplitude to this spectrum, such that the empirically observed negative ODMR spectrum can be simulated and this contribution is displayed in cyan throughout the manuscript.

For the positive ODMR contribution of the $TE_t$ spectrum, we either simulated a randomly oriented molecules-averaged ESR spectrum (for measurements in toluene dispersion) or an in-plane aligned SWCNT film. In the latter case, an orientational distribution function is used in which all SWCNTs long axes are aligned close to parallel to the plane of the film, after which the film was rotated in the external magnetic field. The orientational distribution function uses a Gaussian weight for the angles outside the film plane, with an angular width $\boldsymbol{\sigma_F}$. Afterwards, the contributions to these spectra originating from those angles that provide the negative ODMR contrast were removed from the ESR spectrum, by subtracting the maximum spectrum with negative contrast, without generating any negative signal. This contribution is displayed in dark blue throughout the manuscript.

The spin Hamiltonian parameters, orientational distribution functions and line widths were taken the same for the positive and negative ODMR contrast simulations of the $TE_t$ system, as we assume them to originate from the same triplet system. Since the model nicely approaches all experimental data, this seems to be a valid description.

Fitting of the simulations: Finally, rather than relying on manual adjustments of the parameters displayed in Supplementary Table 2, we chose to optimize numerically the parameters of the spin Hamiltonian, the line widths and the orientational distribution functions. This approach enables a systematic exploration of parameter space, facilitates the identification of correlations between fitting parameters, and provides a more rigorous assessment of parameter uncertainties. To this end we employed a least-squares fitting algorithm based on the *lsqcurvefit* function in Matlab, using a simultaneous fit of all ODMR spectra, while for each ODMR spectrum within an iteration step, the relative amplitudes of these 3 components were determined analytically by linear regression. All parameters numerically optimized through this fit are provided in Supplementary Table 2, with their associated 1-$\boldsymbol{\sigma}$ error bars. To define these error bars, we artificially decreased the degrees of freedom of the fit to twice the number of crossing points of the residuals, to account for systematic deviations of the model with the experimental data and to exclude insignificant data points from the degrees of freedom. The error bars are thus typically larger than obtained by the least-squares optimization performed in MATLAB.

The spin-Hamiltonian parameters were obtained from a global fit of ODMR spectra measured on both randomly-oriented dispersions and frozen dispersions as well as preferentially aligned films. The complementary orientational information provided by these data sets allows orientation-dependent effects to be distinguished from inhomogeneous broadening within the uncertainty of the fit. However, ensemble measurements do not uniquely identify the microscopic origin of the remaining broadening, which may reflect variations in local environment and/or intertube interactions, and the extracted parameters should therefore be regarded as ensemble-averaged values representative of the dominant defect configuration.

**Supplementary Table 2: Simulation Parameters** for defining the spin Hamiltonian parameters, line widths and orientational distribution functions that are used in Easyspin to calculate the associated ESR spectrum. The error bars represent 1-$\boldsymbol{\sigma}$ fit errors. Each of these parameters was numerically optimized by a least-squares algorithm that also takes into account correlations between the parameters.

| | | |
|---|---|---|
| **Triplet System 1 ($TE_t$)** | | |
| $\boldsymbol{g}$-value | $\boldsymbol{g_1}$ | $\mathbf{2.0020 \pm 0.0010}$ |
| Zero-field splitting parameters | $\boldsymbol{D}$ (MHz)<br>$\boldsymbol{E}$ (MHz) | $\mathbf{195.0 \pm 0.4}$<br>$\mathbf{32.4 \pm 1.1}$ |
| Gaussian distribution of ZFS parameters (accounting for inhomogeneous broadening) | $\boldsymbol{\sigma_D}$ (MHz)<br>$\boldsymbol{\sigma_E}$ (MHz) | $16.4 \pm 0.2$<br>$12.5 \pm \mathbf{0.5}$ |
| Intrinsic line width | $\boldsymbol{\sigma_{B,1}}$ (MHz) | $3.78 \pm 0.04$ |
| <u>For films</u>: Gaussian line width of in-plane alignment | $\boldsymbol{\sigma_F}$ (°) | $15.3 \pm \mathbf{0.6}$ |
| <u>For negative component</u>: Gaussian line width of the Z-axis alignment | $\boldsymbol{\sigma_Z}$ (°) | $\mathbf{23.5 \pm 0.6}$ |
| **Triplet System 2 ($TE_p$)** | | |
| $\boldsymbol{g}$-value | $\boldsymbol{g_2}$ | $\mathbf{2.0028 \pm 0.0010}$ |
| Intrinsic line width | $\boldsymbol{\sigma_{B,2}}$ (MHz) | $14.1 \pm 0.2$ |

## Supplementary Note 5: Simulation of Laser Power Dependence

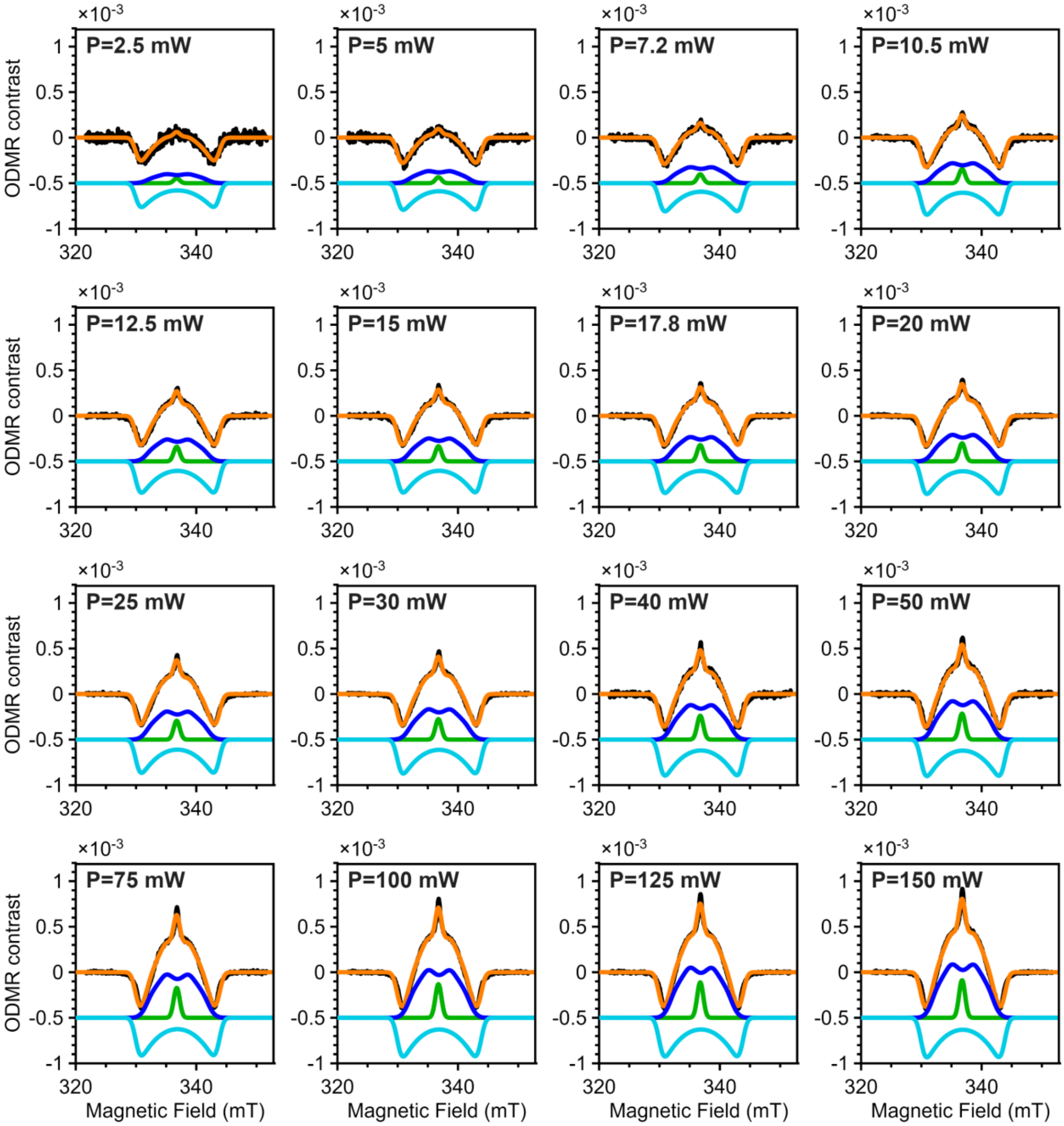

**Supplementary Figure 7**. Simulation of the full-field ODMR spectra in the laser power-dependent measurements including the 3 components as described in Figure 1 and the main text, resulting in Figure 2e-2f. Experimental data are shown in black, fits in orange, and components in green ($TE_p$, $D = 0$), dark blue ($TE_t$, positive signals) and cyan ($TE_t$, negative signals).

## Supplementary Note 6: Simulations of temperature-dependent measurements.

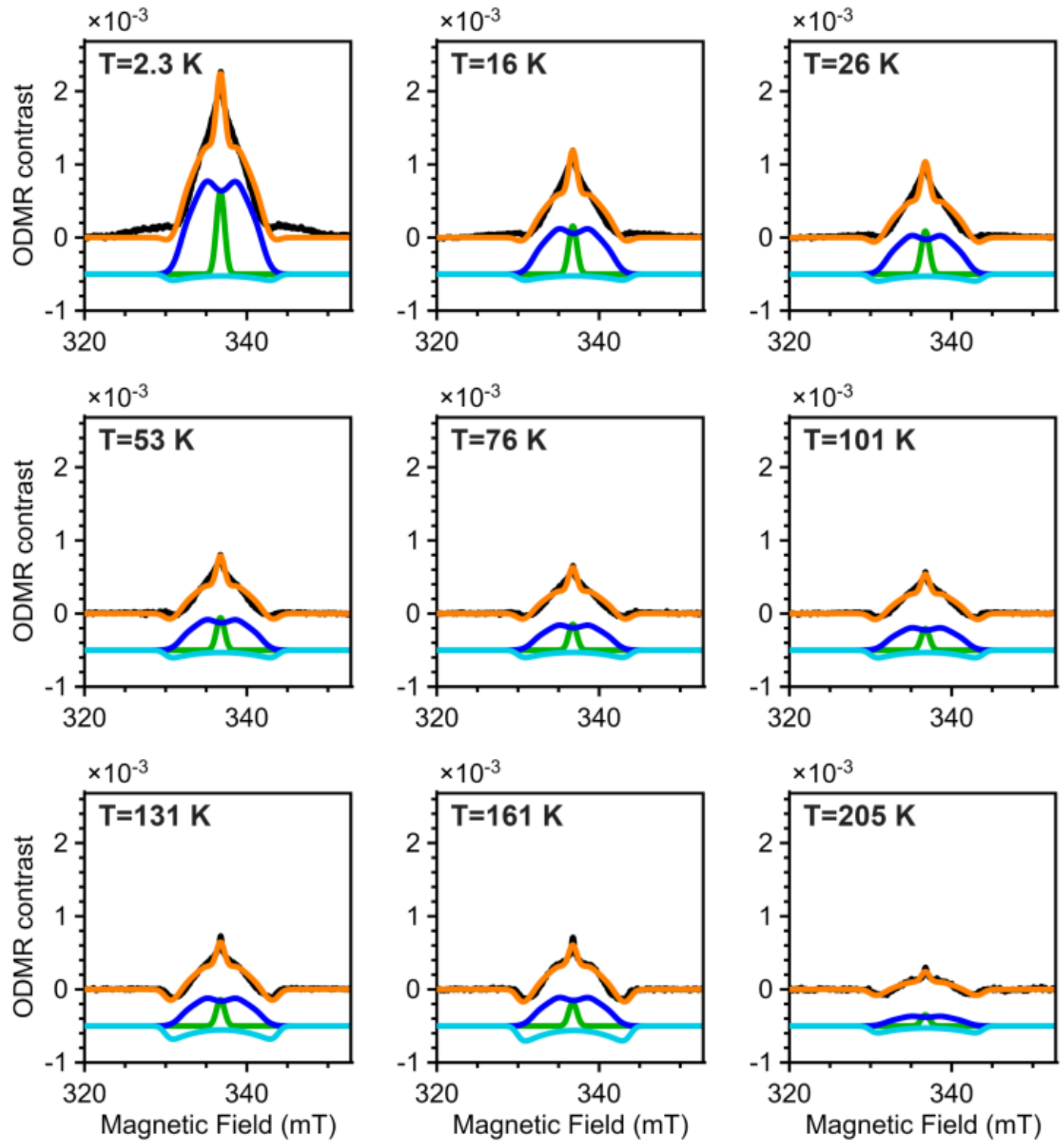

**Supplementary Figure 8**. Simulation of the full-field ODMR spectra in the temperature-dependent measurements from frozen dispersion of (6,5)-I including the 3 components as described in Figure 1 and the main text, resulting in Figure 3d. Experimental data are shown in black, fits in orange, and components in green ($TE_p$, $D = 0$), dark blue ($TE_t$, positive signals) and cyan ($TE_t$, negative signals). Note that at the lowest temperature a third triplet exciton appears, as previously described in reference [2], which was not included in the current fits.

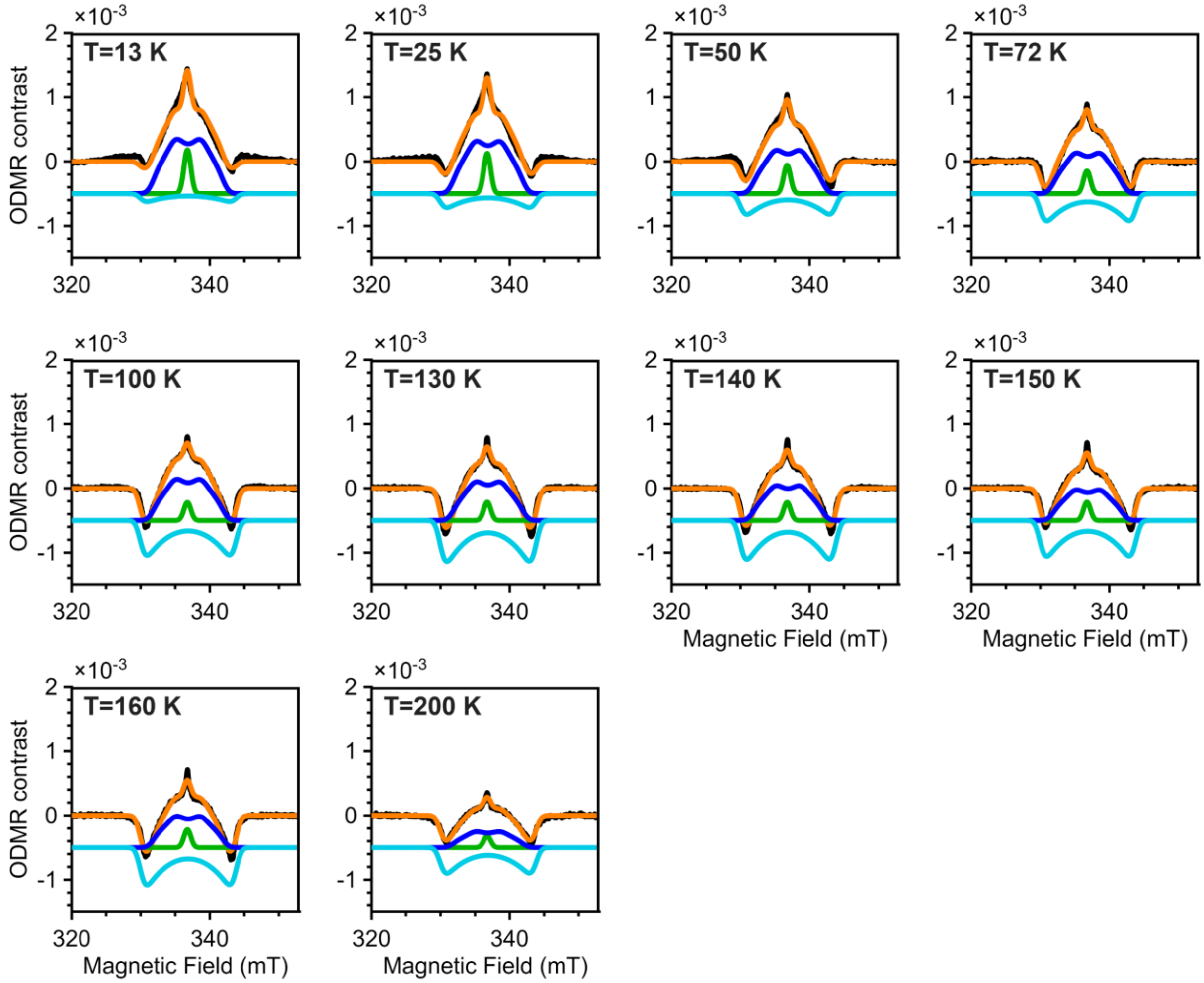

**Supplementary Figure 9**. Simulation of the full-field ODMR spectra in the temperature-dependent measurements from in-plane aligned films of (6,5)-I with magnetic field parallel to the film plane, including the 3 components as described in Figure 1 and the main text, resulting in Figure 3f. Experimental data are shown in black, fits in orange, and components in green ($TE_p$, $D = 0$), dark blue ($TE_t$, positive signals) and cyan ($TE_t$, negative signals).

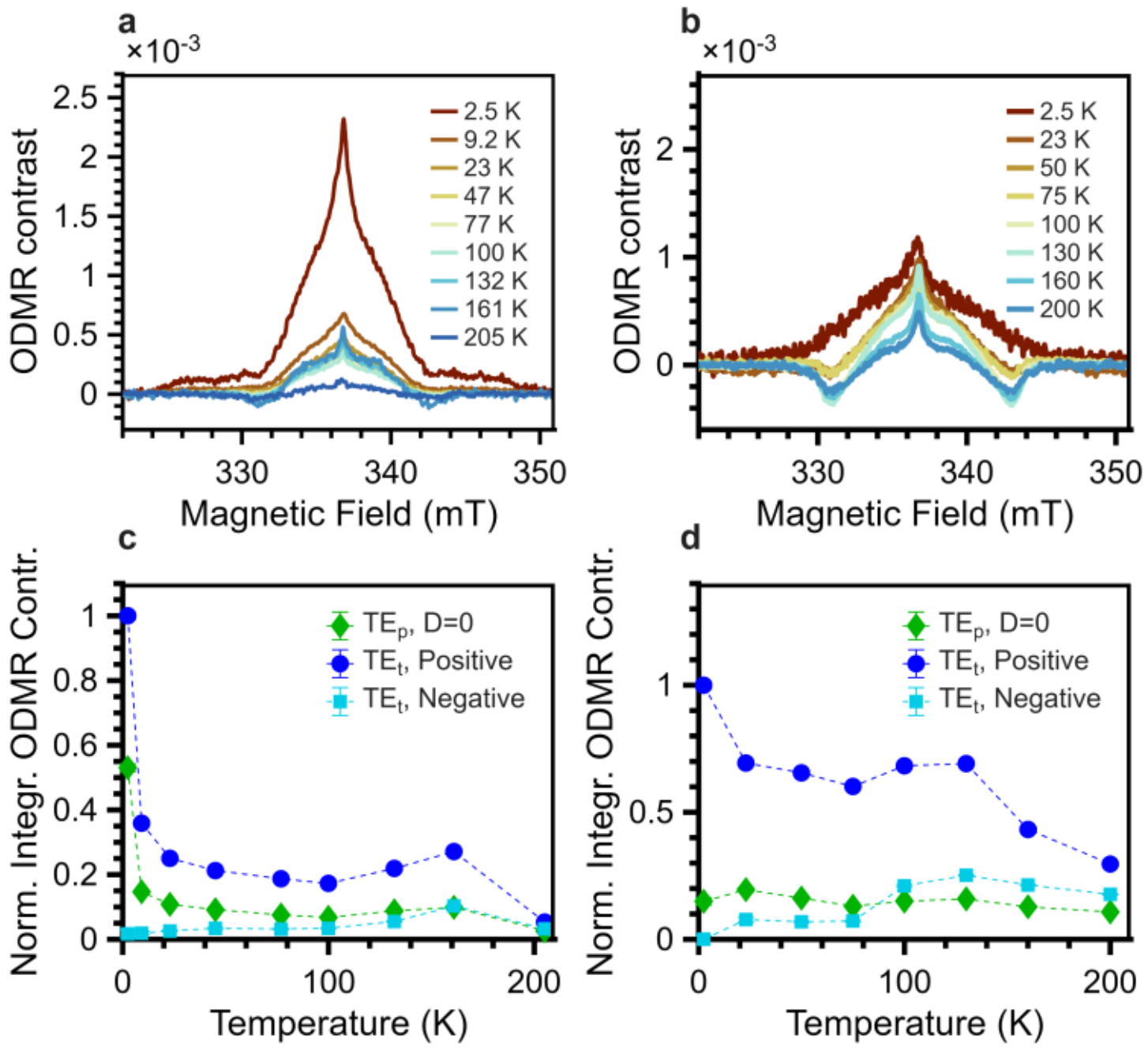

**Supplementary Figure 10**: (a) Full-field ODMR spectra of (6,5)-Br frozen dispersion samples and (b) the (6,5)-Br film sample as a function of temperature. (c) Normalized integrated ODMR contrasts from the three different components with which the ODMR spectra of the frozen dispersion were fitted, with similar color coding as in Fig. 1. Fits are provided in Supplementary Fig. 11. Normalization was done on the highest value of the three curves. (d) Normalized integrated ODMR contrasts from the three different components with which the ODMR spectra of the film samples were fitted, with similar color coding as in Figure 1. Fits are provided in Supplementary Fig. 12. Error bars are 1-$\sigma$ error bars on the fitted amplitudes but are typically too small to visualize. All ODMR measurements were performed with a MW power of 300 mW and a laser power of 100 mW and by passing the emission of the functionalized sample through a 1050 nm long-pass filter.

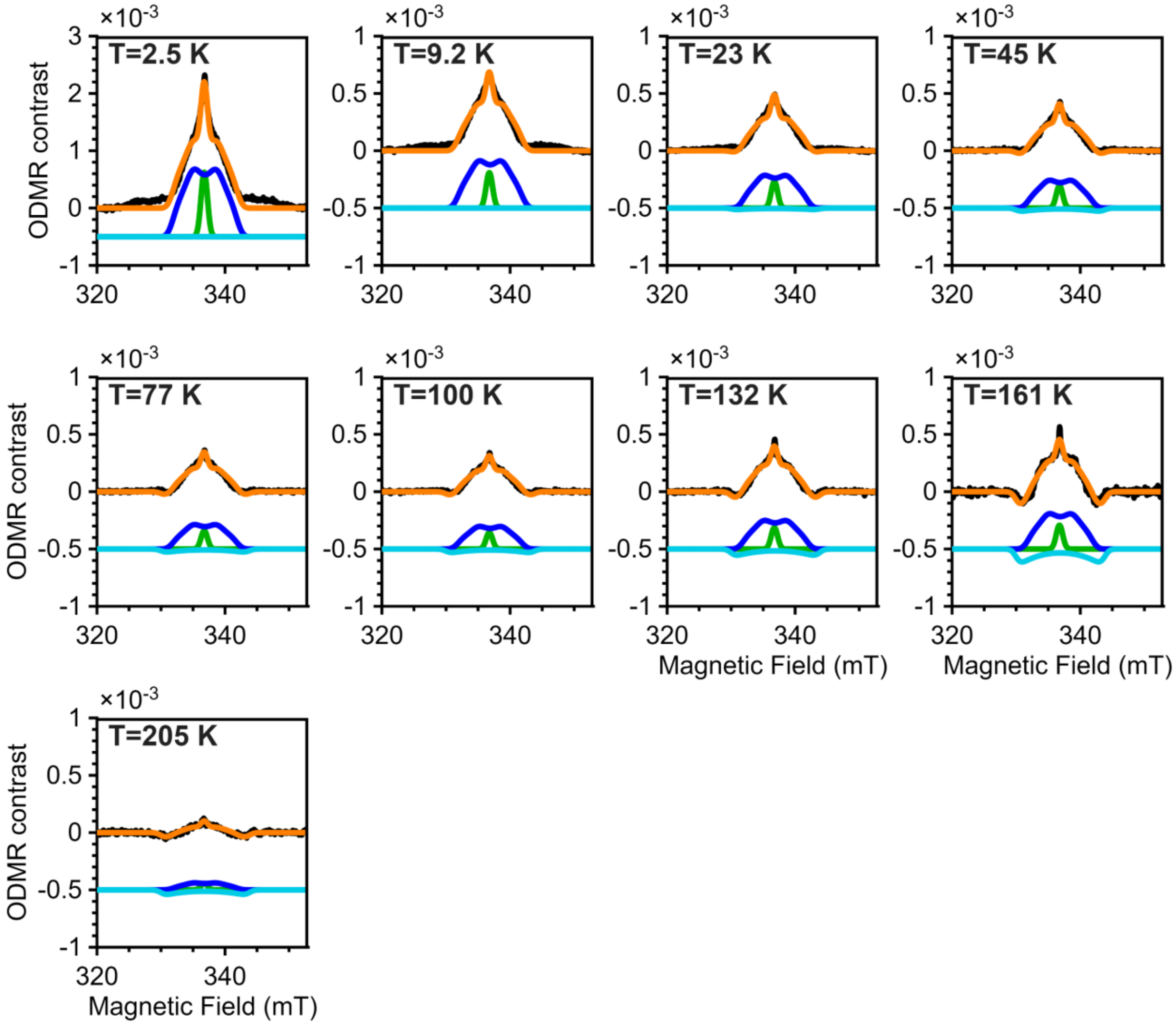

**Supplementary Figure 11**. Simulation of the full-field ODMR spectra in the temperature-dependent measurements from the dispersions of (6,5)-Br, including the 3 components as described in Figure 1 and the main text, resulting in Figure 3f. Experimental data are shown in black, fits in orange, and components in green ($TE_p$, $D = 0$), dark blue ($TE_t$, positive signals) and cyan ($TE_t$, negative signals).

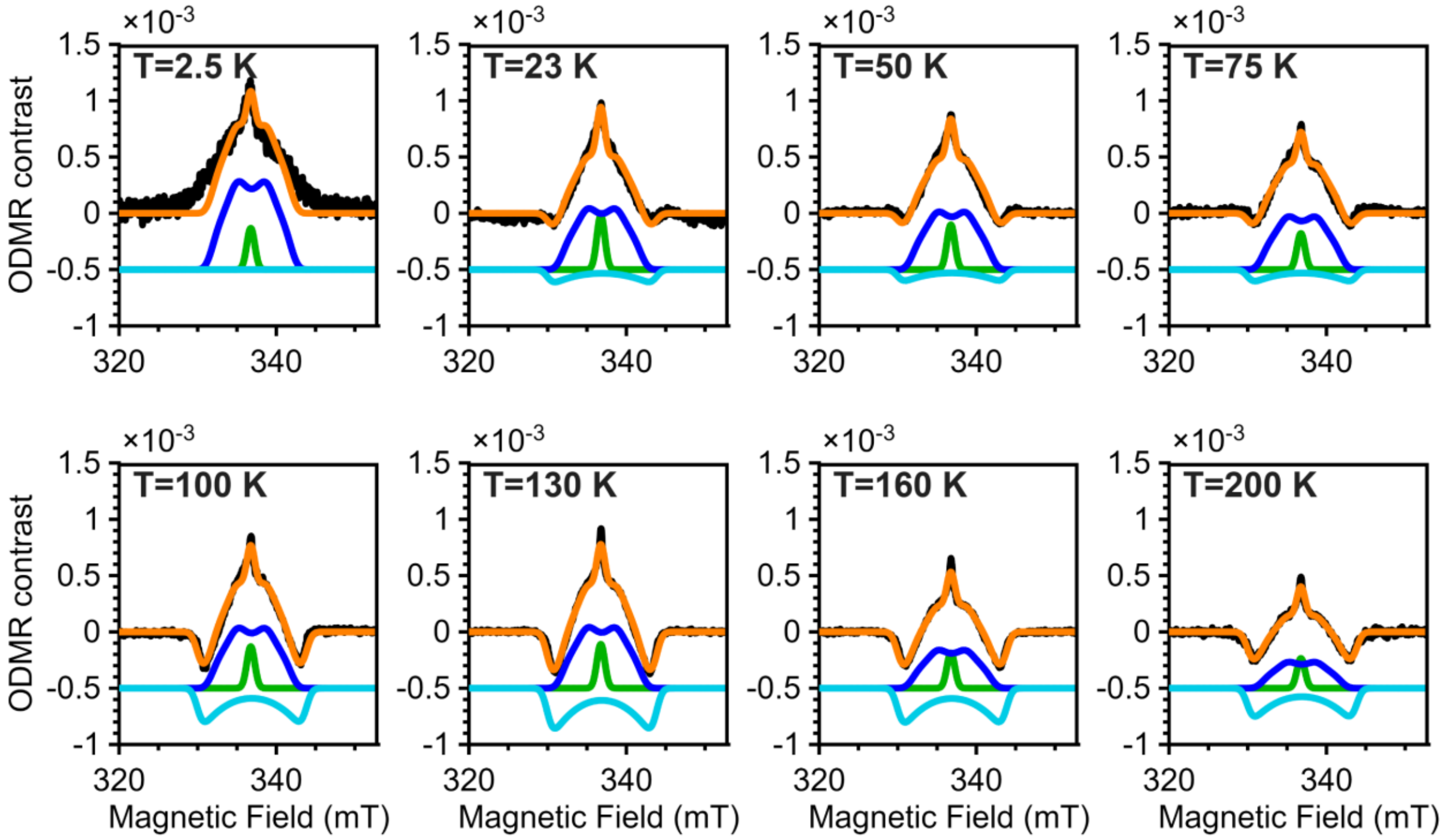

**Supplementary Figure 12**. Simulation of the full-field ODMR spectra in the temperature-dependent measurements from the in-plane aligned films of (6,5)-Br, including the 3 components as described in Figure 1 and the main text, resulting in Figure 3f. Experimental data are shown in black, fits in orange, and components in green ($TE_p$, $D = 0$), dark blue ($TE_t$, positive signals) and cyan ($TE_t$, negative signals).

<u>**Supplementary Note 7: Simulations of angular dependent measurements**</u>

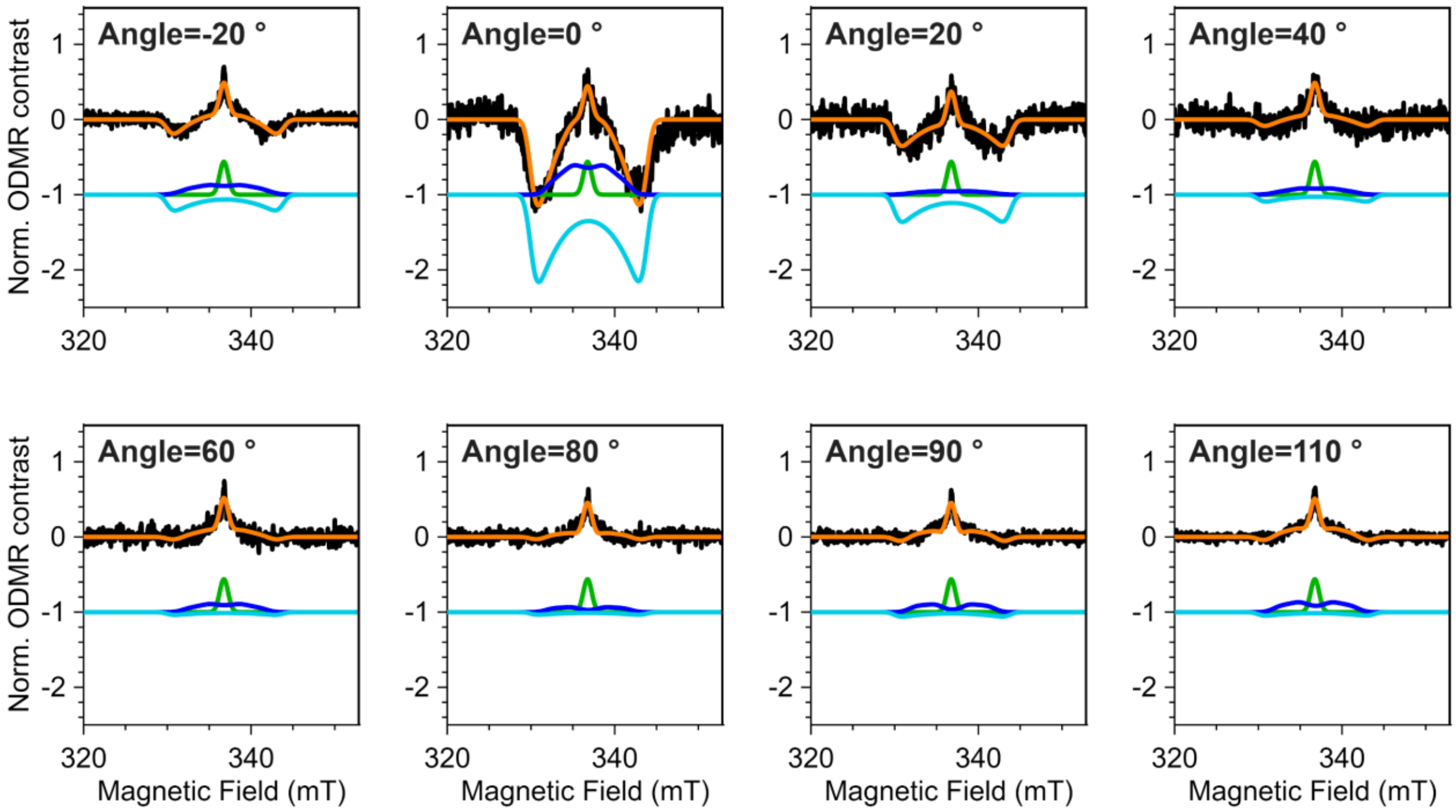

**Supplementary Figure 13**. Simulation of the full-field ODMR spectra in the angular-dependent measurements from in-plane aligned films of (6,5)-I including the 3 components as described in Figure 1 and the main text, resulting in Figure 4. Experimental data are shown in black, fits in orange, and components in green ($TE_p$, $D = 0$), dark blue ($TE_t$, positive signals) and cyan ($TE_t$, negative signals).

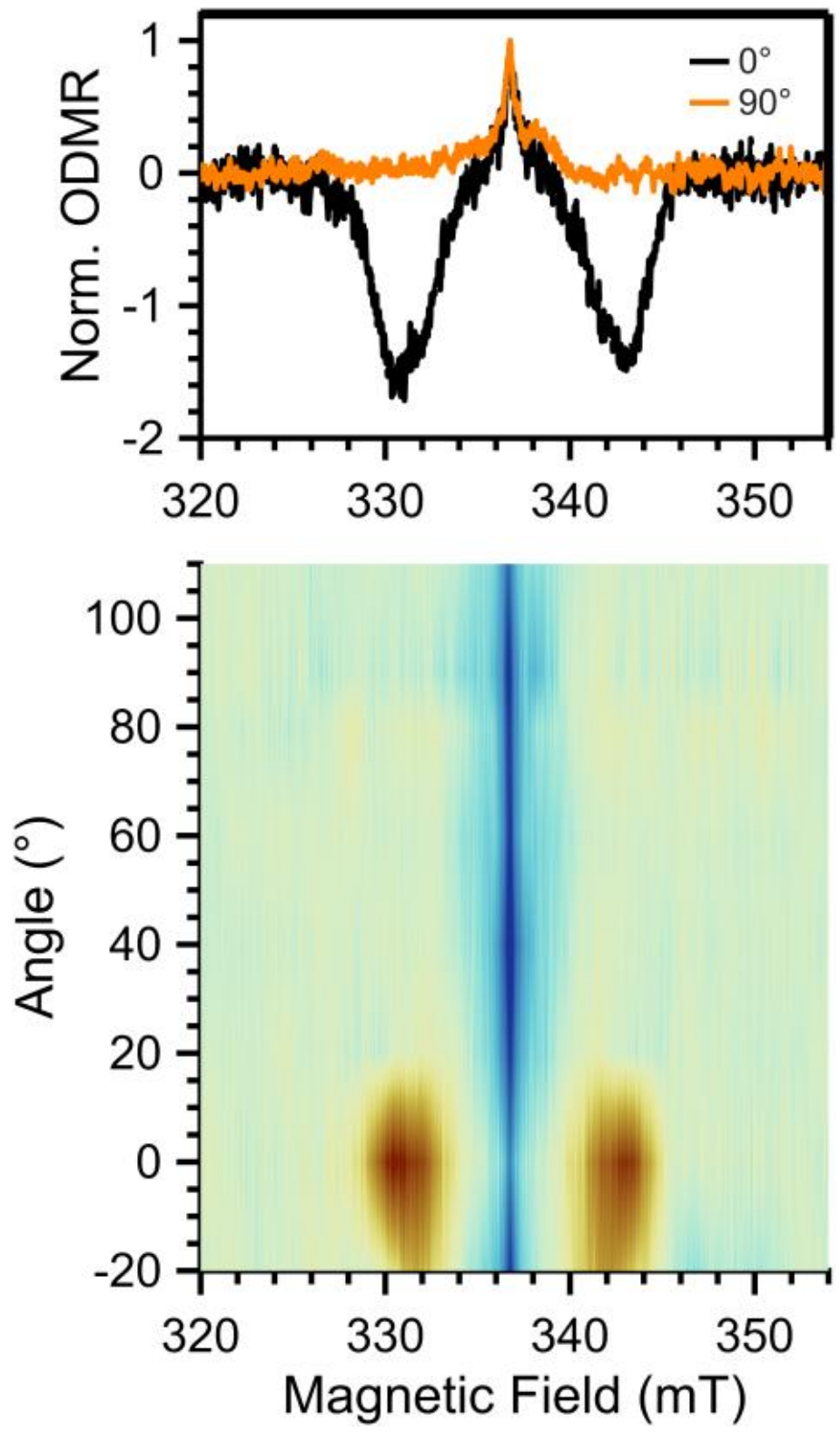

**Supplementary Figure 14**: Angular-dependent ODMR spectra of in-plane aligned (6,5)-I SWCNT films (top) ODMR spectra at room temperature for two discrete angles. All ODMR spectra in this figure are normalized on the fitted amplitude of the $D$=0 peak. (bottom) angular-dependent room-temperature ODMR spectra. The spectra were measured with a MW power of 300 mW, using a long pass filter of 1050 nm, a modulation frequency of 630 Hz, excitation wavelength 561 nm and laser power of 50 mW.

## Supplementary Note 8: Absorption and PL spectra of (6,5)-Cl used for TEMPO sensing

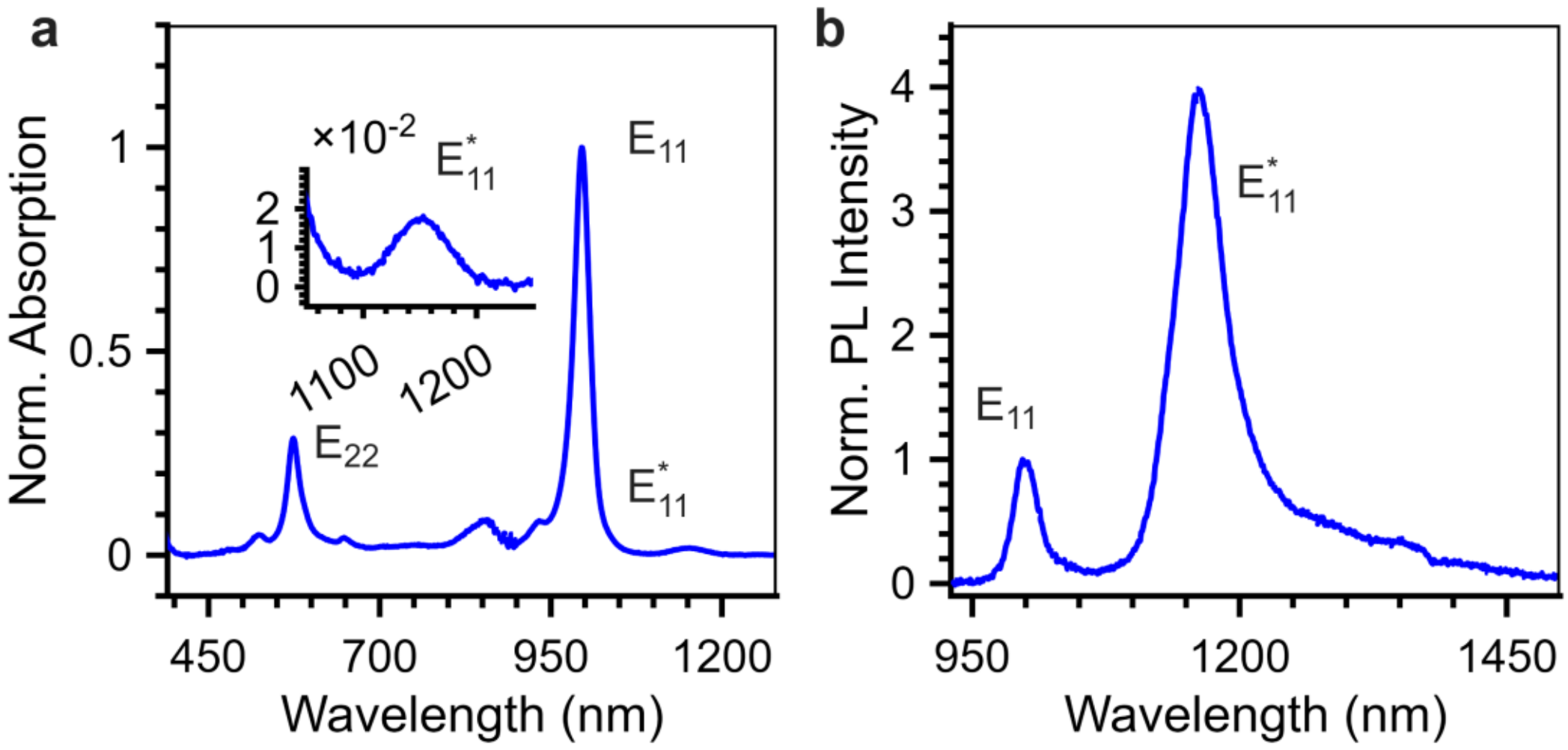

**Supplementary Figure 15**. (a) Normalized absorption spectra of (6,5)-Cl sample used to perform the sensing of TEMPO. The inset zooms in on the $E_{11}^{*}$ absorption band. (b) PL spectra obtained at an excitation wavelength of 570 nm and normalized on the $E_{11}$ emission peak.

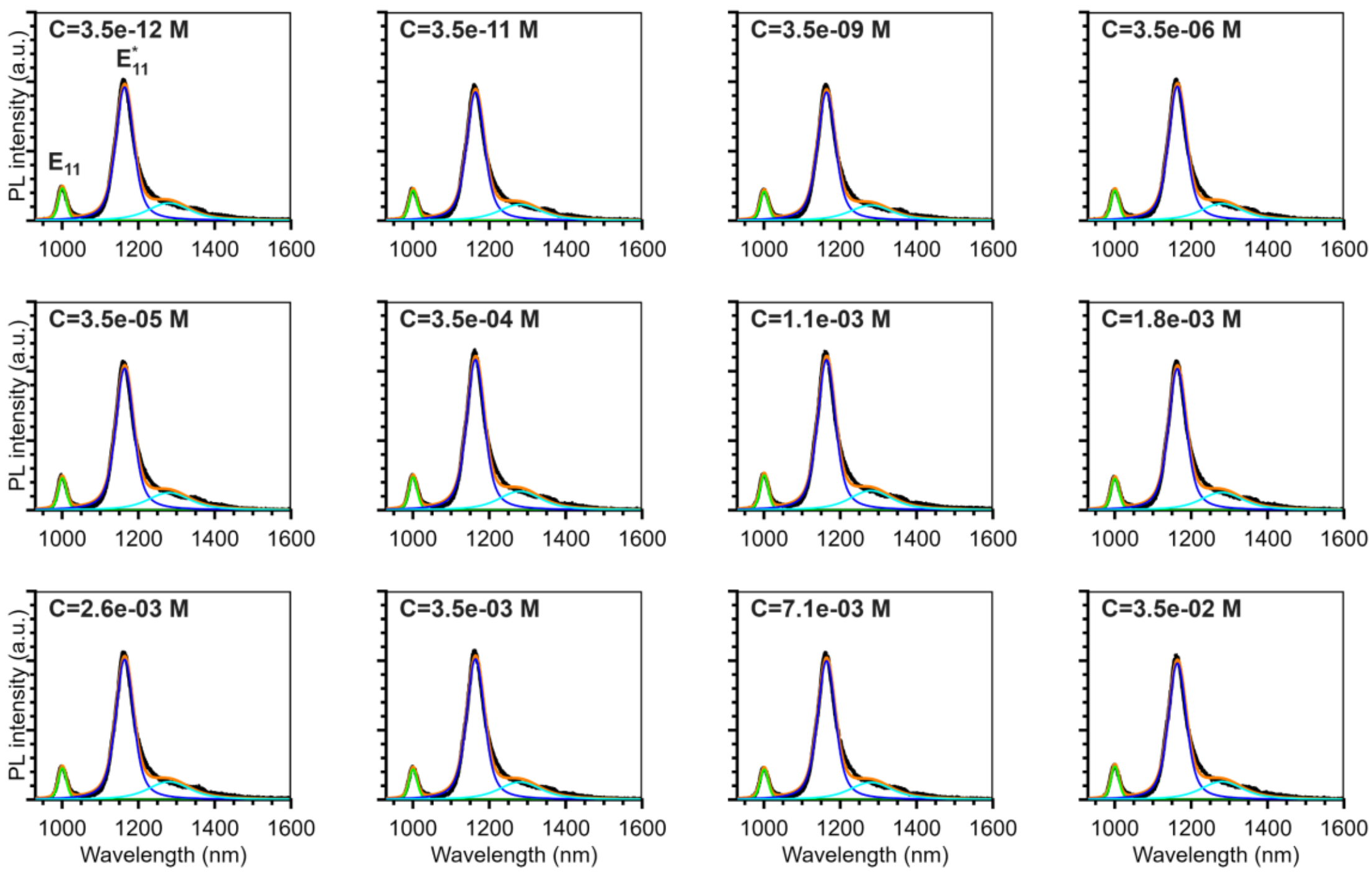

**Supplementary Figure 16:** PL spectra and fits for different concentrations of TEMPO. The data is fitted simultaneously with similar peak positions and line widths, including a Voigt line shape for the $E_{11}$ (in green) and $E_{11}^{*}$ (in blue) emission peak, as well as an additional Voigt line shape for the phonon-side band emissions (in cyan). Intensities of both $E_{11}$ and $E_{11}^{*}$ were extracted from this fit and plotted in Figure 5c. No spectral shifts are observed.

## Supplementary Note 9: Fits of the TEMPO sensing experiment (main text)

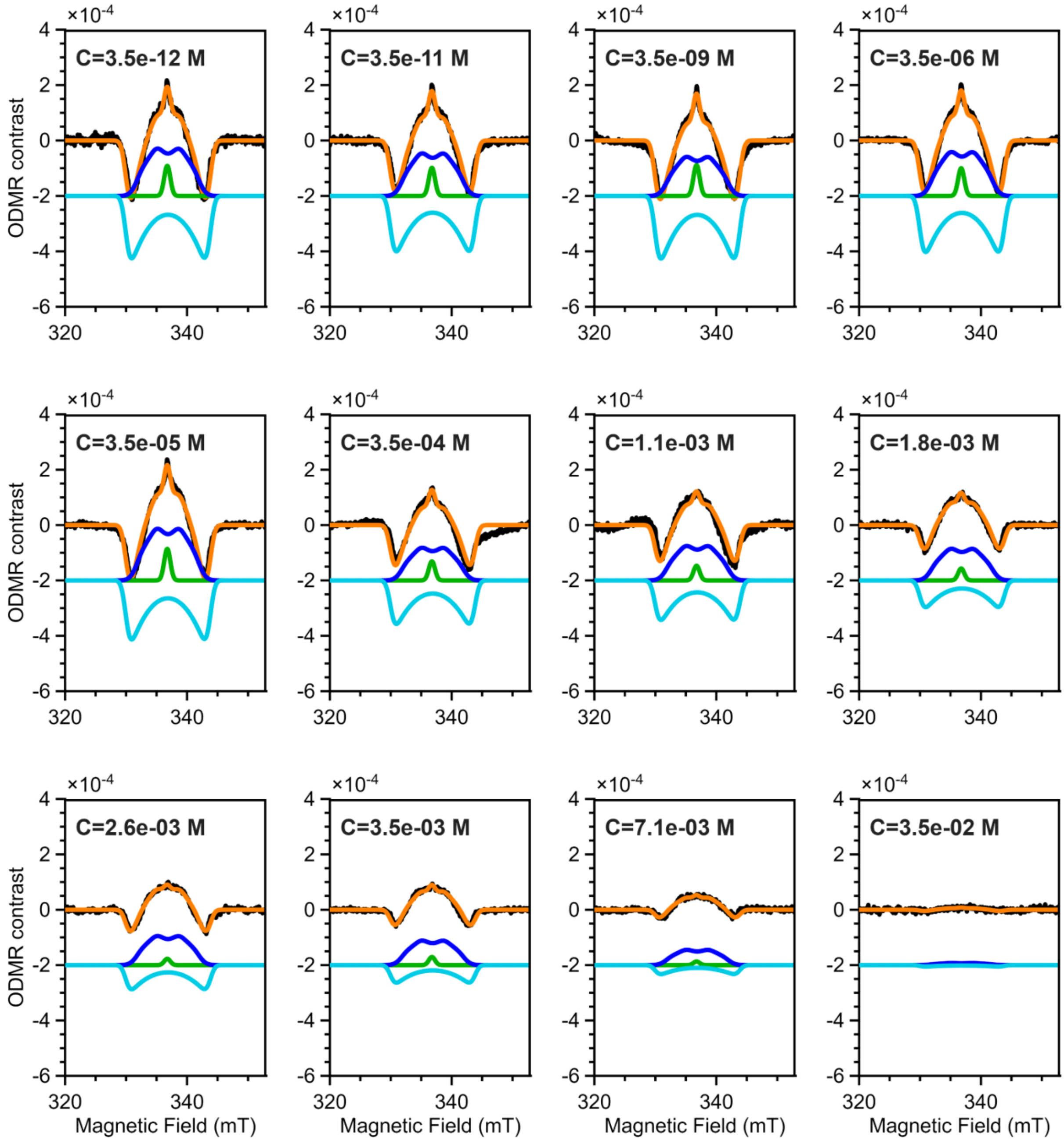

**Supplementary Figure 17**. Simulation of the full-field ODMR spectra in the TEMPO-concentration dependent measurements from in-plane aligned films of (6,5)-Cl including the 3 components as described in Figure 1 and the main text, resulting in Figure 5b. Experimental data are shown in black, fits in orange, and components in green ($TE_p$, $D = 0$), dark blue ($TE_t$, positive signals) and cyan ($TE_t$, negative signals).

## Supplementary Note 10: TEMPO vs. TEMP sensing

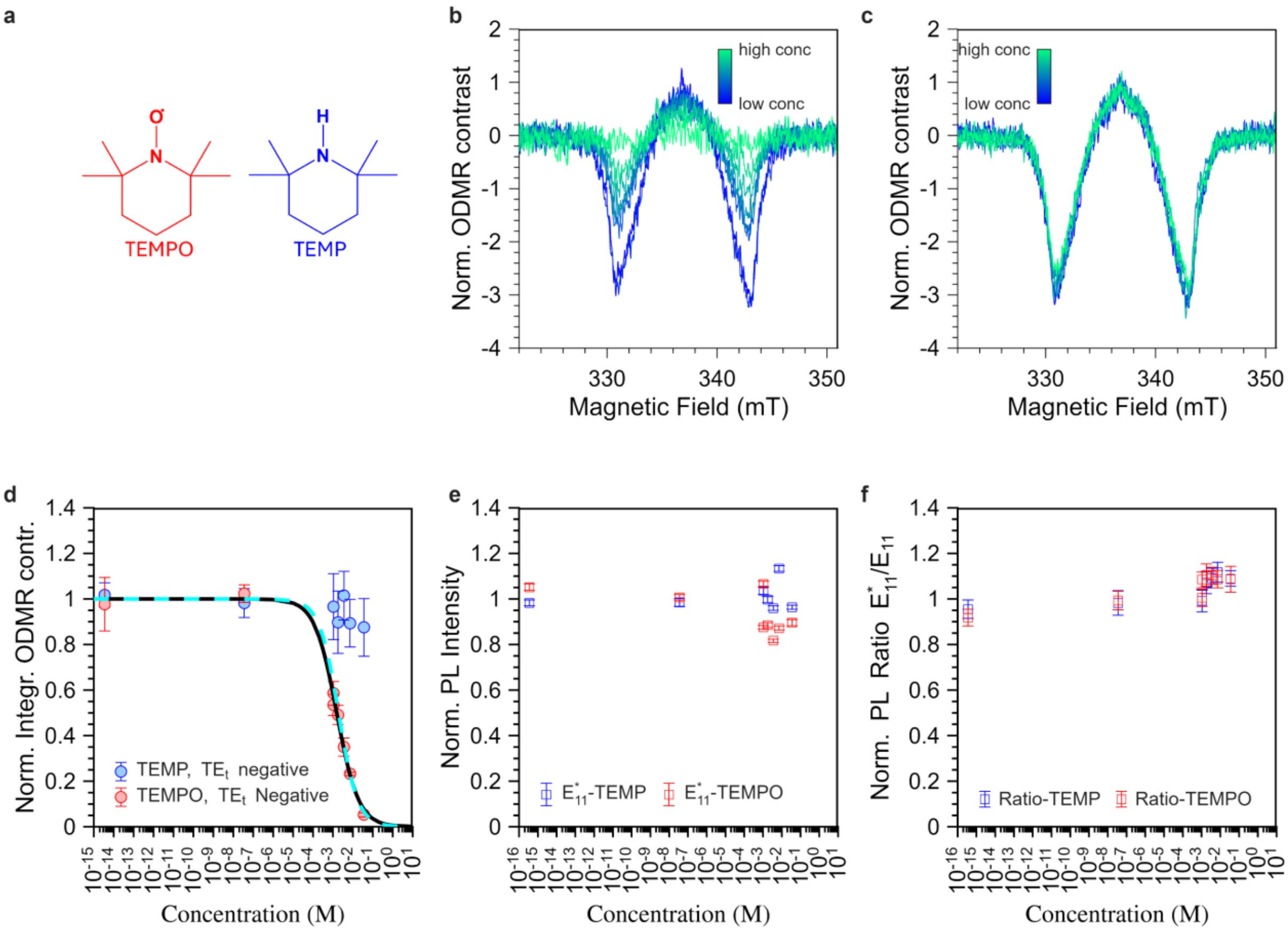

**Supplementary Figure 18.** Additional sensing experiments. (a) Chemical structures of TEMPO and TEMP; (b) RT ODMR spectra as a function of added TEMPO concentration for (6,5)-Br sample. (c) RT ODMR spectra as a function of added TEMP concentration for (6,5)-Br sample, measured as a reference closed-shell molecule. (d) Normalized fitted integrated ODMR contrasts of the TE$_t$ negative component for TEMPO (red circles) and TEMP (blue circles). The TEMPO data were fitted with a Hill model (black solid line) and for comparison also the Hill model of the (6,5)-Br data presented in the main text was added (cyan dashed line). The two sensing experiments are in very close agreement with each other, showing the same sensitivity in spite of the difference between these functional groups. (e) Fitted amplitudes of the $\boldsymbol{E^*_{11}}$ peaks normalized over the mean of all data points, as a function of the TEMPO (red) or TEMP (blue) concentration showing hardly any influence on the SWCNT emission intensity. (f) Ratio of the $\boldsymbol{E_{11}}$ and $\boldsymbol{E^*_{11}}$ peaks normalized over the mean of all data points, as a function of the TEMPO (red) or TEMP (blue) concentration showing hardly any influence on the SWCNT emission intensity.

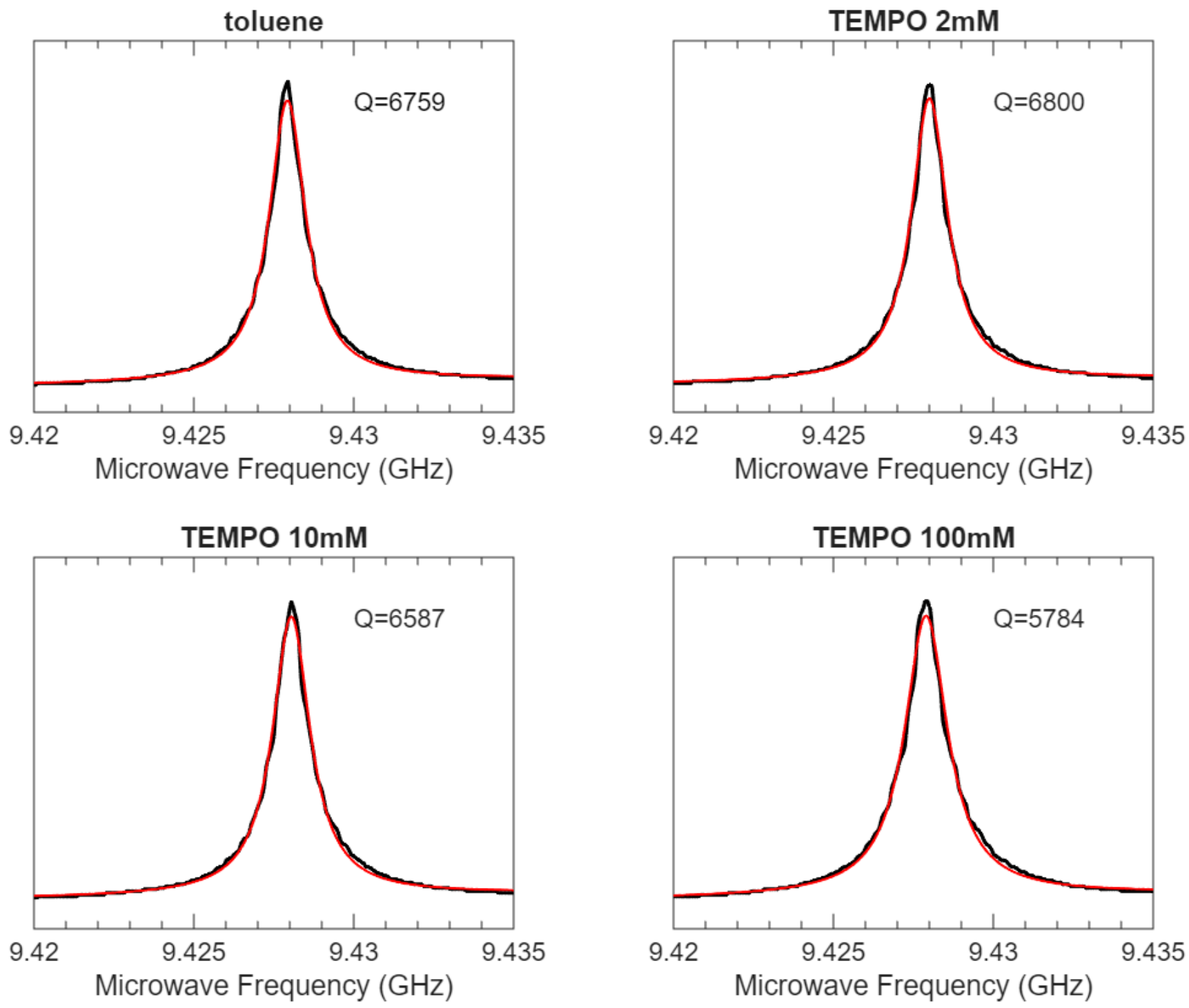

**Supplementary Figure 19**: Inverted and normalized curve of the reflected microwave intensity by the cavity (i.e. the resonance dip of the microwaves) as a function of added TEMPO concentration (black) at a magnetic field in resonance with the central line of TEMPO. Superimposed is a Lorentzian fit (red). The Q-factor was then defined by the equation $Q = \nu/\Delta\nu$ with $\nu$ the microwave frequency and $\Delta\nu$ the fwhm of the dip. Note that at 2 mM, where an overall decrease of 50% of the ODMR intensity is observed in Figure 5 of the main text, the quality factor hardly changes, hence this effect cannot explain the observed change in ODMR intensity.

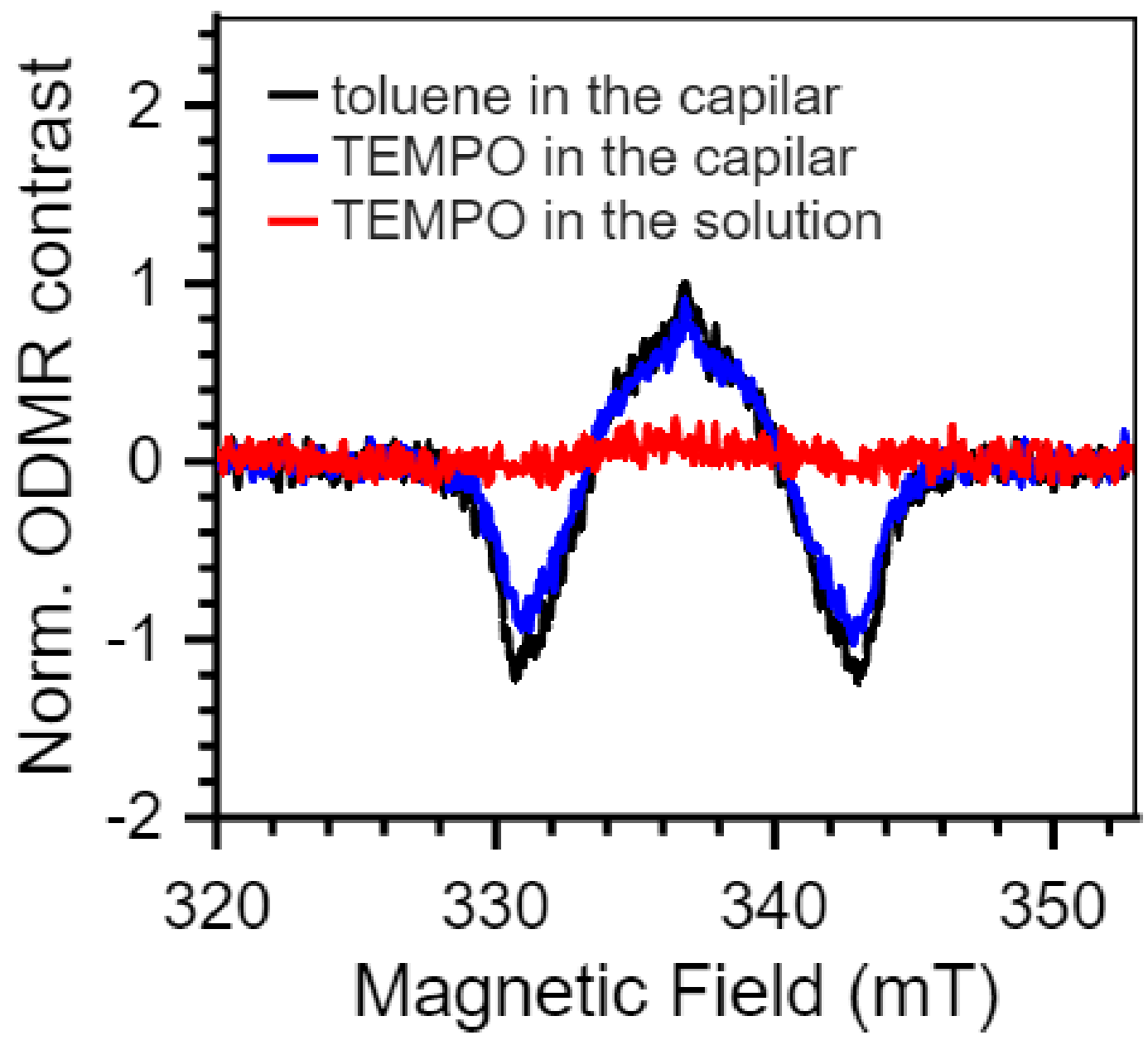

**Supplementary Figure 20**: Separation of cavity-loading effects from direct interactions between TEMPO and the SWCNT triplet states. (black spectrum) Normalized ODMR spectrum measured together with a small, closed capillary with toluene used as a reference. (Blue spectrum) ODMR spectrum, normalized on the black spectrum, taken together with a closed capillary of 100 mM TEMPO in toluene, corresponding to a total concentration (SWCNT sample and capillary combined) of approximately 22 mM, but with the TEMPO radicals physically separated from the SWCNTs so that they cannot interact at the molecular level. (Red spectrum) ODMR spectrum when mixing the same number of spins directly with the SWCNT sample such that a close interaction at the molecular level is possible. Spectra were obtained with a 561 nm laser with power 30 mW, microwave power 360 mW, a modulation frequency of 630 Hz and by passing the emission through a 1050 nm long-pass filter.

## Supplementary Note 11: Influence of the presence of oxygen

To assess the influence of oxygen on the ODMR contrast, we compared a sample measured under ambient conditions with a sample that was first bubbled with argon gas to remove dissolved oxygen. The results, presented in Supplementary Figure 19, show virtually no difference between the two ODMR measurements, indicating that the presence of oxygen does not significantly affect the ODMR contrast.

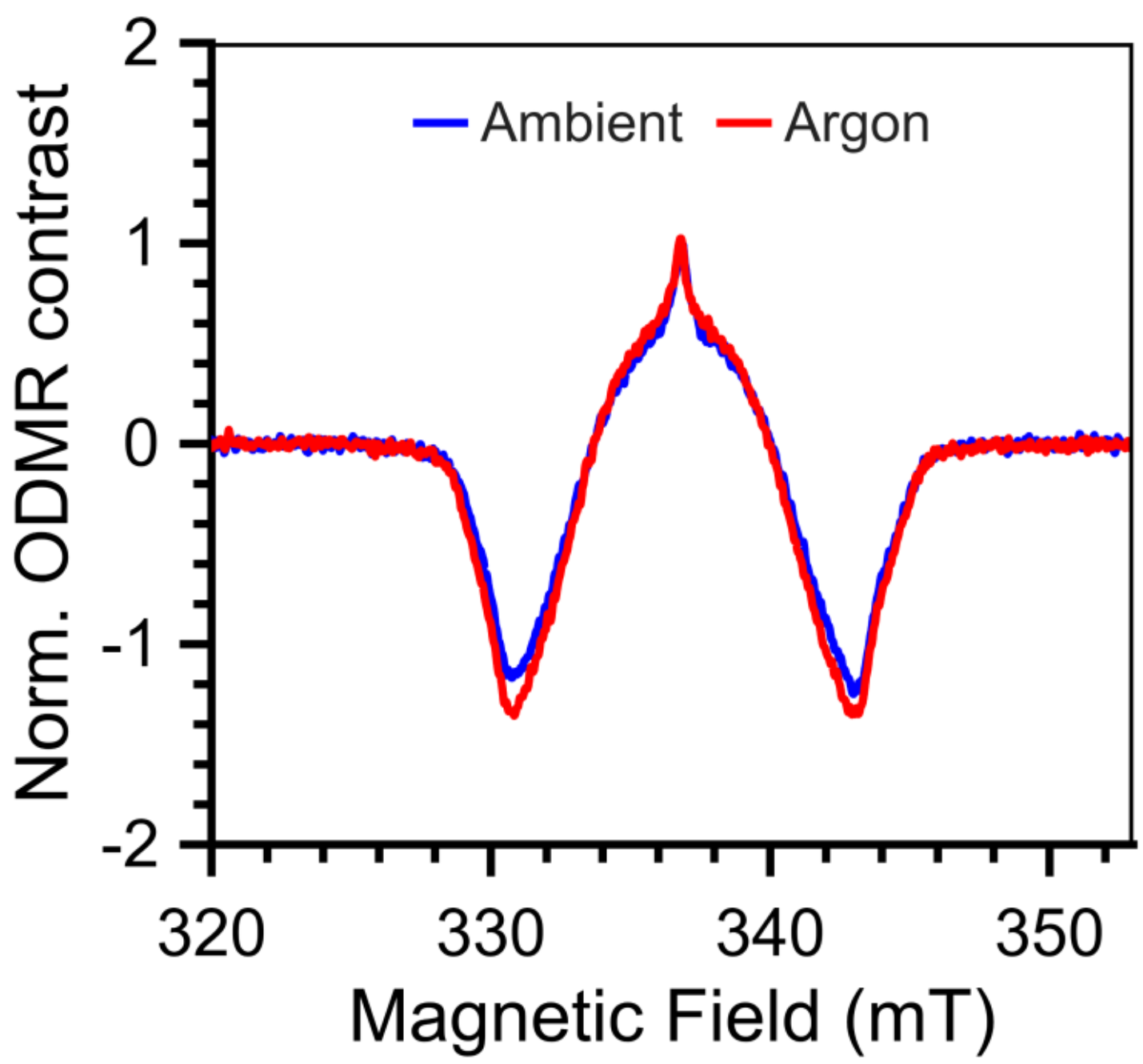

**Supplementary Figure 21**. Full-field ODMR spectra of the same sample either prepared in ambient conditions are after bubbling in Argon for 10 min to remove dissolved oxygen normalized on the maximum of the Ambient spectrum, showing hardly an effect of oxygen on the ODMR contrast

## Supplementary Note 12: Additional references